\documentclass[a4paper,fleqn]{cas-dc}

\usepackage[authoryear]{natbib}
\usepackage{graphicx} 
\usepackage{float}
\usepackage{algorithm}  
\usepackage{algpseudocode}
\usepackage{color}
\usepackage{setspace}

\usepackage{xcolor}

\usepackage{subcaption}
\usepackage{adjustbox}
\usepackage{caption}
\usepackage{lipsum}
\usepackage{csquotes}
\usepackage{bookmark}
\usepackage{xurl}
\usepackage{cleveref}

\usepackage{draftwatermark}
\SetWatermarkText{PREPRINT}
\SetWatermarkScale{0.8}
\SetWatermarkLightness{0.92}

\newcommand{\furl}[1]{\footnote{\url{#1}}}

\def\tsc#1{\csdef{#1}{\textsc{\lowercase{#1}}\xspace}}
\tsc{WGM}
\tsc{QE}
\tsc{EP}
\tsc{PMS}
\tsc{BEC}
\tsc{DE}

\begin{document}
\let\WriteBookmarks\relax
\def\floatpagepagefraction{1}
\def\textpagefraction{.001}
\shorttitle{Virtual Fieldwork in Immersive Environments}
\shortauthors{Bernstetter et al.}

\title[mode=title]{Virtual Fieldwork in Immersive Environments using Game Engines}

\author[1,2]{Armin Bernstetter}[orcid=0000-0003-1603-1699]
\cormark[1]
\ead{abernstetter@geomar.de}
\credit{Writing - Review \& Editing, Conceptualization, Methodology, Software, Investigation, Writing - Original Draft, Visualization}

\author[1]{Tom Kwasnitschka}[orcid=0000-0003-1046-1604]
\credit{Conceptualization, Data Curation, Supervision, Project administration, Funding acquisition, Writing - Review \& Editing}

\author[1]{Jens Karstens}[orcid=0000-0002-9434-2479]
\credit{Data Curation, Writing - Review \& Editing}

\author[1]{Markus Schlüter}
\credit{Software}

\author[2,3]{Isabella Peters}[orcid=0000-0001-5840-0806]
\credit{Conceptualization, Supervision, Funding acquisition}

\address[1]{GEOMAR Helmholtz Centre for Ocean Research Kiel, Kiel, Germany}
\address[2]{Kiel University, Kiel, Germany}
\address[3]{ZBW Leibniz-Information Centre for Economics, Kiel, Germany}

\cortext[1]{Corresponding author}

\nonumnote{\textbf{Frequently used abbreviations:} GIS - geographic information system, HMD - Head-Mounted Display, UE - Unreal Engine, VE - Virtual Environment, VFT - Virtual Fieldwork Tool, VR - Virtual Reality}

\begin{abstract}
    Fieldwork still is the first and foremost source of insight in many disciplines of the geosciences.
    Virtual fieldwork is an approach meant to enable scientists trained in fieldwork to apply these skills to a virtual representation of outcrops that are inaccessible to humans e.g. due to being located on the seafloor.
    For this purpose we develop a virtual fieldwork software in the game engine and 3D creation tool Unreal Engine.
    This software is developed specifically for a large, spatially immersive environment as well as virtual reality using head-mounted displays.
    It contains multiple options for quantitative measurements of visualized 3D model data.
    We visualize three distinct real-world datasets gathered by different photogrammetric and bathymetric methods as use cases and gather initial feedback from domain experts.
\end{abstract}


\maketitle




\section{Introduction}
\label{sec:intro}
Fieldwork still is the first and foremost source of insight in many disciplines of the geosciences.
Inspecting an outcrop in-person to form a mental map or model of the site, corroborated by manual measurements of physical outcrop properties tremendously helps to understand detailed findings in the context of a \enquote{bigger picture}.
This is a professionally well-established thought process that both trains and applies spatial understanding.
Meanwhile, seafloor geologists, as most ocean scientists in general, do not have the luxury of their physical presence at study sites, by nature of the extreme environment of the deep ocean.
They rather rely on data gathered by remotely operated or autonomous underwater vehicles (ROVs/AUVs) or crewed submersibles.
Other research fields that encounter similar problems of too remotely located sites include above all astronomy and planetology, but also volcanology or other fields studying hazardous environments. In this way, our application is similar to the operation of planetary probes or remotely operated space telescopes, where time-critical scientific interests are transmitted remotely, and with a severe time delay.
Besides physical rock and sediment samples, the terrain datasets that put observations into context come in different forms across a range of nested scales, from acoustic methods to seafloor images and video feeds, the latter two of which are then often used to reconstruct highly detailed three-dimensional (3D) models of seafloor outcrops by means of photogrammetric reconstruction \citep{kwasnitschka2013doing,arnaubec2023underwater}.
Based on such 3D models, scientists study the digitally recreated geologic settings on their personal computers using any of the multitude of available geographic information system (GIS) or 3D mesh processing applications (see e.g \cite{escartn2016first}).
Although these allow to view a model from any arbitrary pose, they may easily fail to create a notion of comprehensive situational awareness and realistic spatial perception.
They rely on cognitive immersion rather than spatial, physical immersion in the environment of study employing the full range of bodily perception, even if that spatial immersion may be a simulation.


Research has shown that immersion in virtual environments (VEs) improves the spatial understanding of 3D data such as complex geological models \citep{schuchardt2007benefits}.
Visualizing and viewing models of geological structures in a VE is a natural fit since they are inherently three-dimensional and any classic depiction on paper and even on computer monitors is thus limiting understanding \citep{jones2009integration,caravaca2020digital}.
In VEs, (geological) models can be visualized at real scale, offering better understanding and a realistic sense of the 3D geometry, spatial relationships and distribution of the structure.

In our context, factors that increase the feeling of immersion might include being isolated in a virtual world e.g. when wearing Head-mounted Display (HMD) Virtual Reality (VR) glasses or being surrounded by a spatially immersive environment (see \cref{subsec:bg-spatially}).
Another factor is whether the VE reacts to the user's actions and the user can interact with the virtual world.
This can be achieved by tracking the movement of the head and other body parts or hand-held or wearable components, using one's body in the real world for embodied interactions in the virtual world.
In such an environment, the model can be viewed at real scale as if standing or floating right next to it and embodied interactions can be used to make relevant measurements such as distance, strike and dip, a height profile etc.

Aiming to develop a productive working environment for digital (seafloor) geosciences, our design goals included the following functionality:

\begin{enumerate}
    \item Creation of a sense of presence through spatially immersive visualization and real-time interactive navigation;
    \item High model detail up to realistic eye-limiting resolution, across scales from (sub)centimeter to kilometers;
    \item Real-time quantitative interactions by measuring key structural parameters, e.g. orientation (strike vs dip), distance or size
\end{enumerate}

Leveraging the recent major advances in computer graphics enabled by the gaming industry, we chose to develop our workflow based on the Unreal Engine\furl{https://www.unrealengine.com} (UE), a powerful virtual 3D environment originally intended for game development.
Here we introduce an application built in UE version 5.3 intended to examine nested acoustic large-area and photogrammetric close-up seafloor terrain models employing both HMDs and our spatially immersive visualization laboratory, the ARENA2\furl{https://www.geomar.de/en/arena} (\cref{fig:arena1}). We call this approach a virtual fieldwork tool (VFT) \citep{geomar60840}.

Due to the research area of our institute, our use cases are focussed on marine geology. The software presented in this work, however, is able to visualize any digital outcrop model regardless of its location on the Earth and thus providing a tool for many different geosciences.


\section{Related Work and Background}
\label{sec:background}
Our concept links subdisciplines of the geosciences, computer graphics and human-computer interaction (HCI) studies.
Below, we provide an overview of related previous research and introduce relevant aspects of the multifaceted, currently practiced bathymetrical workflow that motivates our work.

\subsection{Bathymetric Seafloor Mapping}
\label{subsec:bg-photogrammetry}

As water is largely impermeable to far reaching electromagnetic fields, (e.g., radar, infrared light) \citep{mobley1995optical}, we rely on mechanical methods – primarily acoustics –  to map the seafloor.
These methods come with inherent trade-offs in resolution versus range.
Therefore, the commonly practiced way of surveying the seafloor employs a stack of several methods across an overlapping cascade of scales.
It ranges from ship-based multibeam swath echo sounders (with a typical resolution of 1\% of the water depth, i.e., tens of meters and yielding quasi-textures based on the strength of the acoustic return signal) to close-range echo sounders, sub bottom profilers and sonars deployed from deep-towed sleds, ROVs or AUVs, achieving resolutions down to single meters or even centimeters.
The majority of such acoustic terrain models has a 2.5D data structure, with the elevation represented as an extrusion of a plane with no double values possible (forming a Digital Elevation Model, DEM \citep{guth2021digital}).
Thus, steep, (sub-) vertical or even overhanging structures are poorly imaged, or not at all, despite the observation that these typically form the geoscientifically most relevant outcrops due to their lack of contemporary sediment cover, granting access to the vertical sequence of deposits.

To mend this geometrical shortcoming, but also to further increase geometric resolution into the millimeter scale, and to provide a color texture in the familiar human visible light range, photogrammetric methods have been added to the stack of survey methods throughout the last decade \citep{kwasnitschka2013doing,arnaubec2023underwater}.
These require the surveying of outcrops from close distances of typically 2-8m using subsea cameras and lighting arrays.
This may yield realistic terrain models that fully serve the human visual senses equivalent to on-land, in-person field studies.
Depending on the methods employed, the final terrain model takes the form of a point cloud or a mesh, optionally with a texture.

\subsection{Spatially Immersive Displays}
\label{subsec:bg-spatially}
We define a spatially immersive display environment as one that physically covers the full human field of view, and ideally more.
The CAVE \citep{cruzneira1992cave} is one of the early such examples, featuring a 4-sided room in which visitors could stand in and immerse themselves in the environment projected onto the walls and the floor.
Motion tracking systems enable the VE to react to user movement, enabling embodied interactions, while motion parallax and stereoscopic rendering increase the plasticity of the simulation.
In the context of geoscientific collaborative work, they have been successfully used for visualizing LiDAR data \citep{kreylos2008immersive,hsieh2011visual}, atmospheric data \citep{helbig2014concept}, and interactive visualizations of multi-scale geological models
\citep{schuchardt2007benefits,jones2009integration,hyde2018immersive}.

The original CAVE concept was continuously extended and modernized over the years.
Facilities with the goal of enabling collaborative sensemaking through spatially immersive visualization can be found all over the world, for example the CAVE2 at UIC \citep{febretti2013cave2}, Brown University's YURT \citep{kenyon2014design}, the AixCAVE at RWTH Aachen \citep{kuhlen2014vadis} or the Allosphere at UC Santa Barbara \citep{hoellerer2007allosphere}, in addition to a long standing commercial interest throughout the oil and gas industry \citep{evans2002future}.
At GEOMAR we operate the ARENA2 (see \cref{fig:arena}), an architecture based on the concept of digital projection domes \citep{kwasnitschka2023spatially}.

\begin{figure}
    \centering
    \includegraphics[width=\ifdim\linewidth<15cm\linewidth\else0.8\linewidth\fi]{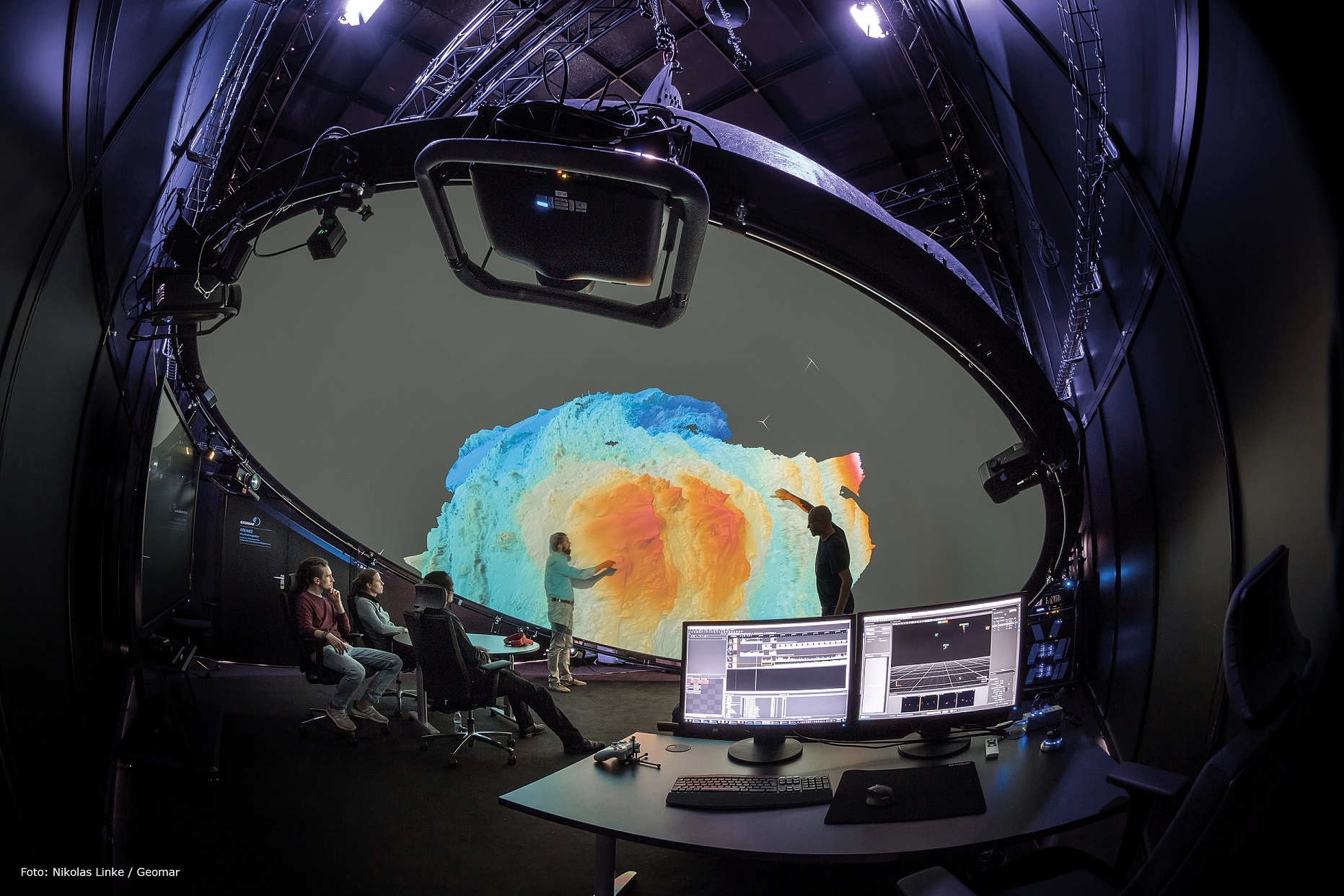}
    \caption{Interactively analyzing a seafloor bathymetry 3D model inside the ARENA2 at GEOMAR Helmholtz Centre for Ocean Research Kiel.}
    \label{fig:arena1}
\end{figure}

With the emergence of on-set virtual production technology in cinematography in recent years, becoming popular since its use in the series "The Mandalorian" \citep{purtill2023virtual}, the technology that is available for spatial immersion has improved.
One of the drivers of this development may have been the COVID-19 pandemic during which filmmakers were not able to travel to and film on location and had to find solutions \citep{purtill2023virtual}, similar to how ocean researchers are not able to travel to the deep seas in person.
The ecosystem used for many of these virtual production stages and setups is based on UE.

\subsection{Virtual Geoscientific Fieldwork and Collaborative Immersive Analytics}
\label{subsec:bg-virtualfieldwork}
Immersive analytics is a research field that makes use of VEs to facilitate interactive analysis of visualized data \citep{chandler2015immersive, skarbez2019immersive, fonnet2021survey}.
Research in this area dates back to the early 2000s \citep{vandam2000immersive,kreylos2006enabling, kreylos2008immersive} but with the advent of increasingly available and affordable VR and HMD technology there have been more possibilities to engage in this field.
Some work in immersive analytics targets more traditional forms of data visualization like graphs \citep{cordeil2017imaxes} whereas others utilize immersion and VEs for improved spatial understanding of complex 3D structures \citep{schuchardt2007benefits}, e.g. medical and molecular data \citep{kuk2023state}, astrophysical data \citep{bock2020openspace}, or geoscience data \citep{jones2009integration, klippel2019transforming, klippel2020value, seers2022virtual}.

Virtual geoscientific fieldwork has frequently been used to support or simulate real-world fieldwork either for training students or for actual scientific sensemaking \citep{klippel2019transforming, klippel2020value}.
Partly driven by the Covid-19 pandemic, \cite{bursztyn2021fostering} for example developed a tool that trains geology students in learning the geological strike and dip measurement convention in a VE, and improving their spatial visualization skills, an important part of geosciences \citep{titus2009characterizing}.
%
With the Virtual Reality Geological Studio (VRGS)\furl{https://www.vrgeoscience.com/}, \cite{hodgetts2007integrating} has built a powerful albeit closed-source toolbox for field geologists to visualize and analyze 3D outcrop models either on a desktop or in HMD VR.
Its set of interactive analytical tools is a holistic portfolio of tools required for (immersive) virtual fieldwork applications.

Many approaches to immersive analytics are utilizing game engines for scientific work and visualization which is a dynamically evolving field of research \citep{friese2008using, reina2020moving}.
\cite{krueger2024engines} describe several requirements for and their experience with using UE as a tool for scientific immersive visualization in their AixCave \citep{kuhlen2014vadis} leading up to their decision to fully switch from custom built software architectures to UE.

\cite{gerloni2018immersive} employ the Unity game engine for the exploration of geological environments in immersive virtual reality, \cite{harrap2019engine} use it to visualize a spatial simulation of rockfalls, and \cite{zhao2019harnessing} visualize the point cloud of a volcanic vent in Unity including several interaction methods.

\cite{bonali2024geavr} present GeaVR, a tool programmed in Unity for immersive VR using HMDs that has grown over the years and includes a multitude of use-cases and tools for the exploration of and interaction with geological structures and sites. Through user evaluations, they were successful in showing its usefulness especially for the education of geoscience students \citep{bonali2022academics,wright2023student}.
\cite{caravaca2020digital} leave the realms of the Earth and utilize Unity to enable virtually visiting a digital model of an outcrop located on Mars.
\cite{billant2019performing} on the other hand are targeting the seafloor with their open-source software Minerve, visualizing a 3D model of the Roseau Fault \citep{escartn2016first}, and implementing fundamental geological measurements, which are also supported in our system.

\cite{cerfontaine2016immersive} and \cite{wang2020geovreality} make use of UE for the immersive visualization of geophysical data and \cite{huo2021efficient} use UE to visualize large-scale oblique photogrammetry models.
In the context of the ARENA2 as our main use case, UE is without alternative due to its high-performance virtual production ecosystem.

What is sometimes challenging for immersive analytics, however, is the aspect of collaborative work \citep{cordeil2017immersive,benk2022assessing}.
Especially HMDs cause a certain degree of isolation from the surroundings which reduces interaction with people outside the VE to sound, unless a multi-player environment has been implemented.
The shared physical presence of a field party in an outcrop, their shared work, and discussions on site are an important aspect of collaborative geoscientific work which can be simulated in spatially immersive environments.

\cite{metois2021oceanic} apply Minerve \citep{billant2019performing}, which supports multiple users, for teaching and evaluate its benefit for virtual fieldwork.
Similarly, \cite{caravaca2020digital} also support multiple users in HMD VR with their system.
In our case, we make use of the ARENA2 providing co-located collaboration between multiple users in the same physical space.


\section{The ARENA2: Spatially Immersive Environment for Geoscience}
\label{sec:arena}

The ARENA2 is a multi-projection dome situated at GEOMAR Helmholtz Centre for Ocean Research Kiel. The evolution of spatially immersive visualization domes at GEOMAR is laid out by \cite{kwasnitschka2023spatially}.

\subsection{Architecture and Hardware}
\label{subsec:a2-architecture}
\begin{figure}[pos=ht]
    \centering
    \begin{minipage}[t]{\linewidth}
        \centering
        \adjustbox{valign=t}{\includegraphics[width=\ifdim\linewidth<15cm\linewidth\else0.8\linewidth\fi]{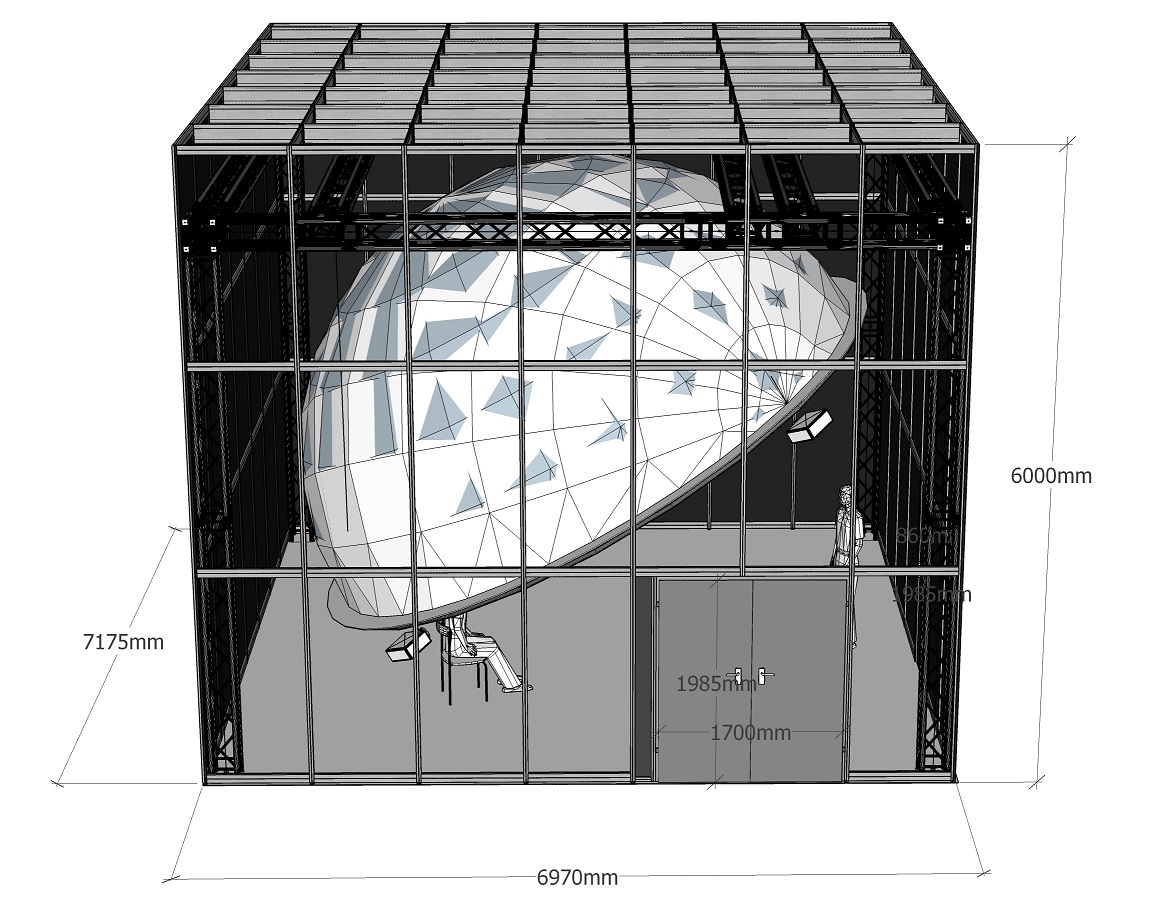}}
        \subcaption{A cut-away sketch of the ARENA2. Shown is the roughly $7 m \times 7 m \times 6 m$ enclosure which is fully opaque in reality. Also visible is the scaffolding on which the free-hanging dome structure is suspended at a 21° angle.}
        \label{fig:arenacutaway}
    \end{minipage}%
    \vspace{1em} 
    \begin{minipage}[t]{\linewidth}
        \centering
        \adjustbox{valign=t}{\includegraphics[width=\ifdim\linewidth<15cm\linewidth\else0.8\linewidth\fi]{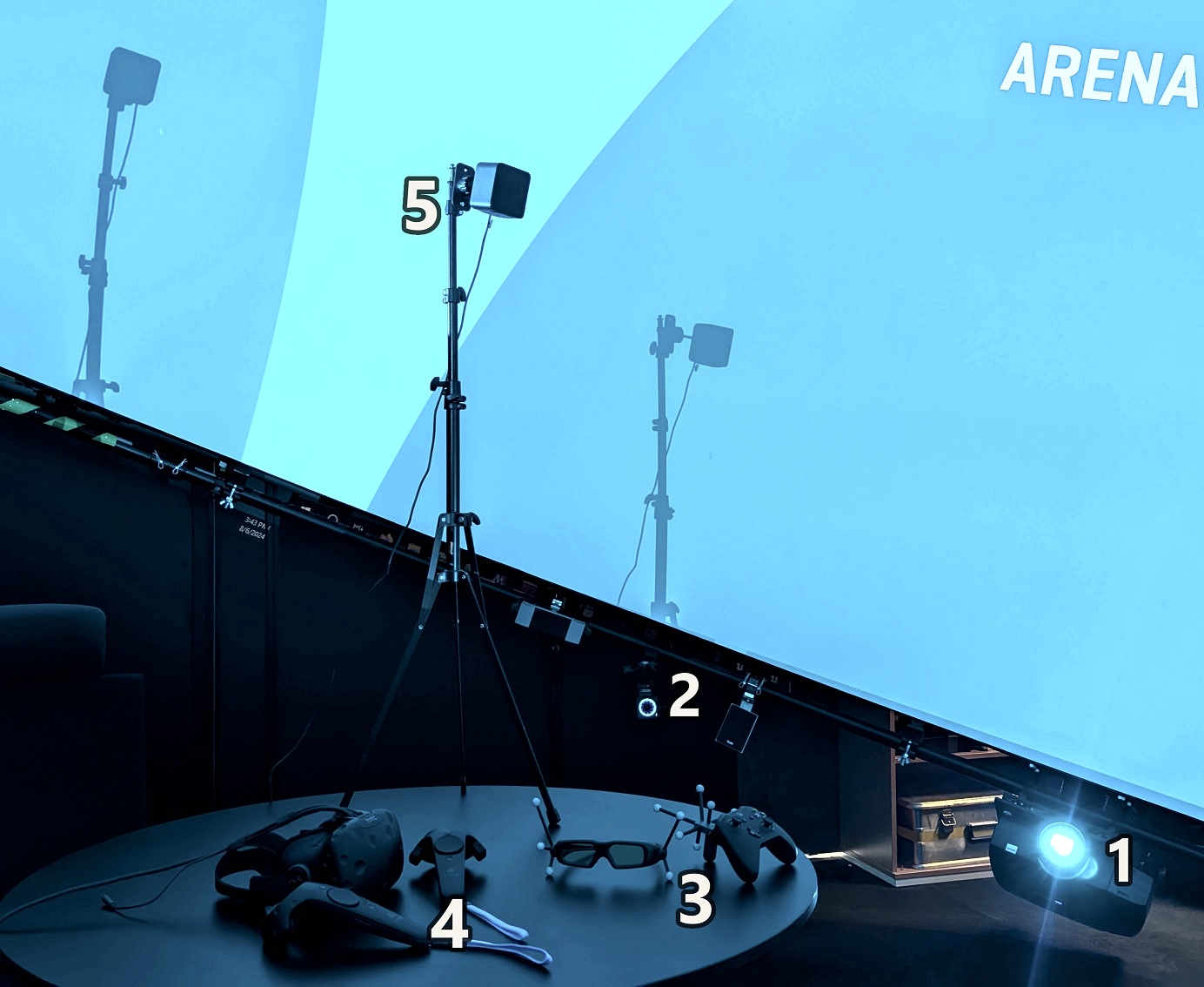}}
        \subcaption{An overview of several of the devices we are using in the ARENA2. Visible in the background is the inside surface of the projection dome. \textbf{1:} One of the five BARCO F50 WQXGA projectors. \textbf{2:} One of the four OptiTrack motion tracking cameras, recognizable by the blue glowing ring. \textbf{3:} Stereo 3D shutter glasses and an Xbox controller equipped with tracking markers. \textbf{4:} An HTC Vive VR headset with the accompanying controllers. \textbf{5:} One of two HTC Vive base stations (\enquote{lighthouse}) used to track the position of the Vive headset and controllers.}
        \label{fig:arena_devices}
    \end{minipage}

    \caption{Images showing a sketch of the ARENA2 (\cref{fig:arenacutaway}) and a selection of the devices we are using (\cref{fig:arena_devices}).}
    \label{fig:arena}
\end{figure}

The ARENA2 dome has a diameter of 6 meters and a tilt angle of 21° to make it possible to display content at eye level as well as above users. \Cref{fig:arenacutaway} shows a sketch of the architecture and \cref{fig:arena1} shows the inside of the ARENA2.

It is equipped with five WQXGA (2560x1600) projectors that also allow stereoscopy, a 5.1 surround sound system and a set of four OptiTrack\furl{https://optitrack.com/} cameras for motion tracking.
A Windows computer cluster of five nodes is used for real-time interactive applications.
The dome projection screen is mathematically described by calibration files in the VESA MPCDI standard\furl{https://vesa.org/vesa-standards/}.
It specifies the warping (i.e., the geometric reprojection of planar imagery onto the curved dome surface) and blending of color and brightness among overlapping projector frustums (i.e., the viewing sectors of each channel on the dome) to create a seamless surface.

\subsection{Unreal Engine in Multi-Display Cluster Setups}
\label{subsec:a2-unreal}

Profiting from the technical advances in virtual production, we use UE technology developed for distributed multi-display and multi-projection setups.
The concept of virtual production stages has emerged as an alternative to filming in green screen studios. It uses virtual environments pre-built in UE to reduce workload in post-production and increase immersion for actors who now have visual references such as a horizon instead of a green wall where the background is added afterwards.
A virtual camera frustum in the VE responds to the perspective and movement of a motion tracked real camera \citep{purtill2023virtual}.

The main technology that enables the rendering of UE content to a cluster of multiple computer nodes and displays is the nDisplay plugin \citep{dalkian2019ndisplay}.
Using our OptiTrack setup, a user's viewpoint is connected to active stereo shutter glasses providing head-tracking and 3D in the virtual environment. Two gamepads equipped with tracking markers are used as input devices for (collaborative) embodied interaction  (See also \cref{fig:arena_devices}).


\section{Implementing a Collaborative Virtual Fieldwork Tool in a Game Engine}
\label{sec:unreal}
The Unreal Engine is a game engine and 3D creation tool enabling programmers and artists to develop games and other content both in C++ as well as a powerful visual scripting environment called \enquote{Blueprints}\furl{https://dev.epicgames.com/documentation/en-us/unreal-engine/introduction-to-blueprints-visual-scripting-in-unreal-engine}.

\begin{figure*}
    \centering
    \includegraphics[width=0.75\linewidth]{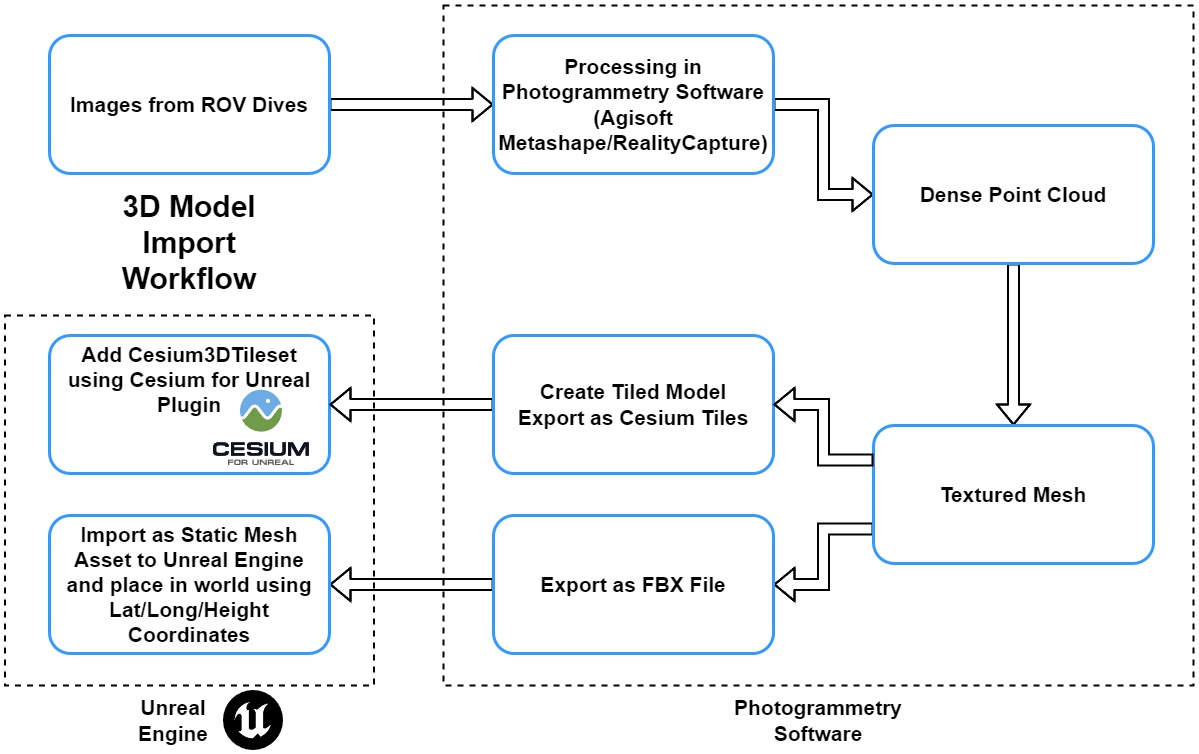}
    \caption{Our photogrammetry preprocessing workflow for adding models to our Unreal Engine VFT. Images are processed in the respective photogrammetry software and a mesh is created which receives a texture calculated also from the images. This textured mesh can then either be exported directly as Cesium tileset or to an FBX file and imported as static mesh asset to UE.}
    \label{fig:photogrammetry}
\end{figure*}

\subsection{Preprocessing}
\label{subsec:vf-prep}

A georeferenced context is a prerequisite for any application aiming at exploration of real-world geospatial data visualized as 3D models.
Our system uses Cesium \citep{cesium_website} and the Cesium for Unreal plugin \citep{cesium_unreal_repo} for creating a georeferenced representation of the Earth.
Cesium is an ecosystem and platform for 3D geospatial data, also offering open source software that enables the creation of georeferenced environments in various different applications and frameworks such as UE.
We are using precisely geolocated coordinates (longitude/latitude) which is possible because Cesium and UE itself support using double-precision floating-point numbers.
In Cesium for Unreal, coordinates can be accessed both as geodetic as well as Earth-centered, Earth-fixed coordinates.
These are translated by Cesium to the UE coordinate system where one internal unit of measurement equals exactly 1 centimeter i.e. 100 UE units are 1 meter.
Due to the double-precision floating-point representation these UE units can have decimal values of up to 15 decimal digits.
The coordinates and any measurements are thus calculated in a way to ensure the best possible accuracy.

\Cref{fig:photogrammetry} shows our workflow (see also \cref{subsec:bg-photogrammetry}) leading up to being able to load 3D models in UE.
Once exported as FBX file, 3D models can be imported into UE as static meshes where they can be placed into our georeferenced Cesium world using their geodetic coordinates.
During import to UE, a few further optimizations are advisable:
One should ensure that the imported meshes have fine-grained collision that enables querying the model at any point of its surface.
Furthermore, a custom material has to be applied to the imported mesh to be able to use the clipping box (see \Cref{subsec:vf-toolbox}).

Another, more streamlined workflow is to export photogrammetric models directly in the Cesium 3D tiles format\footnote{\url{https://github.com/CesiumGS/3d-tiles}} \citep{cozzi20233dtiles} with which the models are automatically placed at their correct coordinates on the Cesium Earth ellipsoid.
This approach removes several steps that might be prone to human error or decimal point inaccuracies stating the mesh coordinates.

Our Cesium tilesets are always being rendered on their highest level of detail, providing a fine-grained collision mesh which is important for accurate measurements.

Cesium also allows a runtime data-loading workflow since Cesium3DTilesets are streamed into the UE world and not imported as UE proprietary asset. Given the (file) url, a new Cesium3DTileset can be added during runtime without having to manipulate its location, material, or collision.


%

\subsection{Interaction}
\label{subsec:vf-interaction}

\begin{figure}
    \centering
    \begin{minipage}[t]{\linewidth}
        \centering
        \adjustbox{valign=t}{\includegraphics[width=\ifdim\linewidth<15cm\linewidth\else0.8\linewidth\fi]{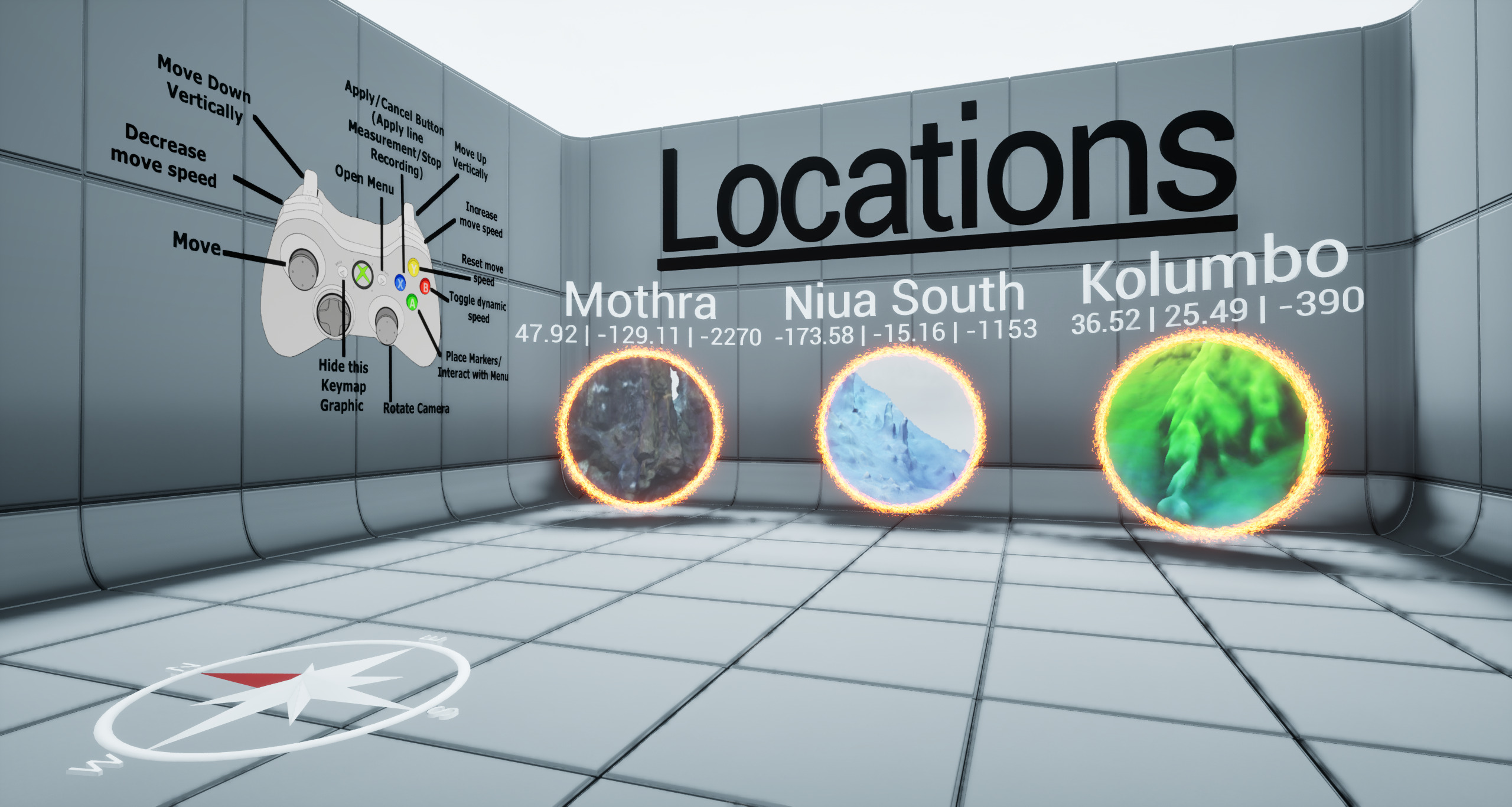}}
        \subcaption{The starting level in which a user is placed initially when starting the VFT. The \enquote{portals} labeled with the name of the location teleport the user to the respective model. The compass widget visible in the lower left part of the image can be hidden.}
        \label{fig:unrealstart}
    \end{minipage}%
    \vspace{1em} 

    \begin{minipage}[t]{\linewidth}
        \centering
        \adjustbox{valign=t}{\includegraphics[width=\ifdim\linewidth<15cm\linewidth\else0.8\linewidth\fi]{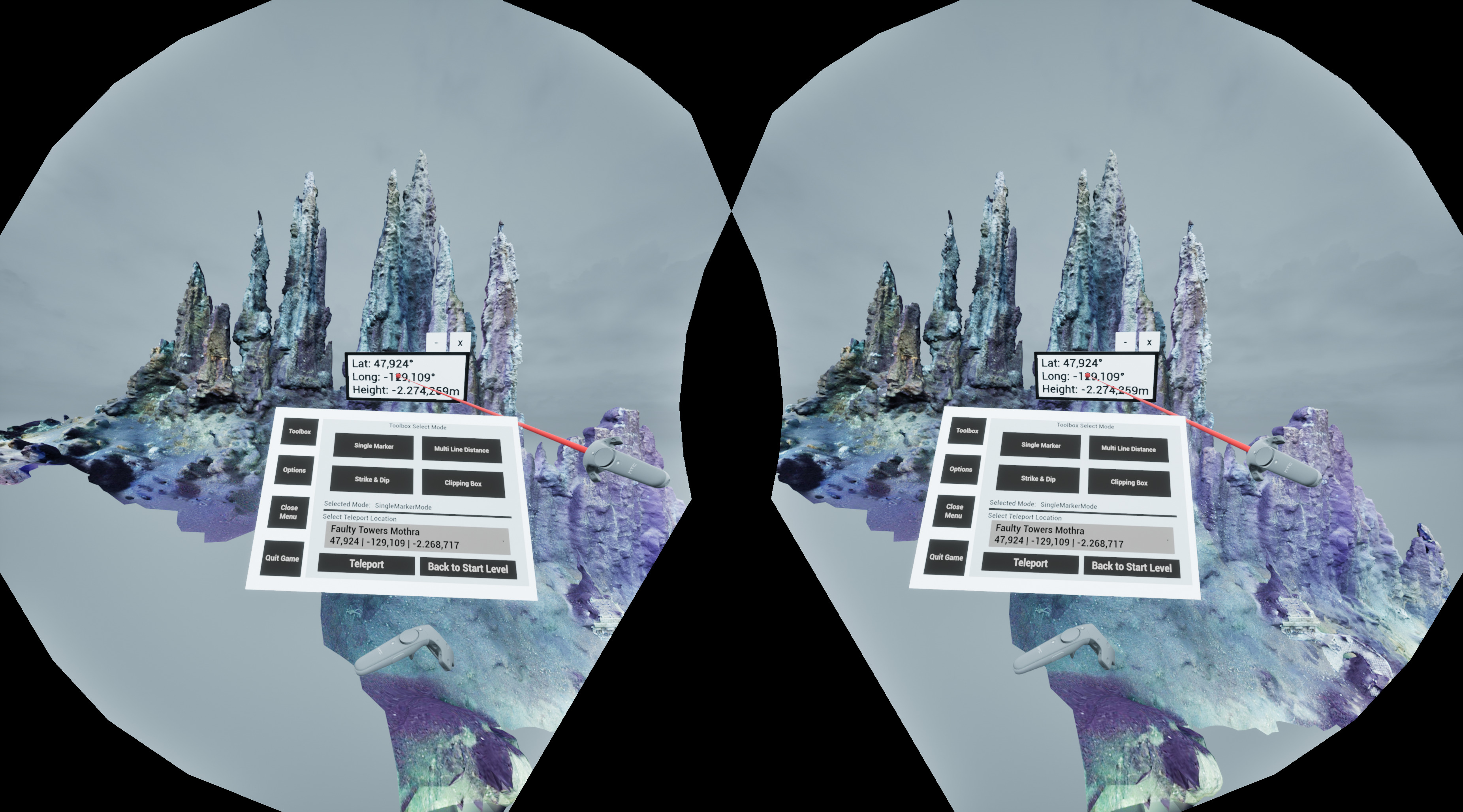}}
        \subcaption{A screenshot of the HMD VR setting showing what is being rendered for both eyes. The image also shows the menu widget attached to the left HTC Vive controller and the \enquote{pointer ray} emanating from the right controller.}
        \label{fig:unrealvr}
    \end{minipage}

    \caption{Screenshots from the VFT showcasing the user experience}
    \label{fig:unreal}
\end{figure}

Analogous to many existing HMD VR applications, embodied interaction with the virtual world is implemented by means of pointer rays attached to the controllers. A 3D menu widget allows options to be selected by targeting them with the ray and pressing a button (see \cref{fig:unrealvr}).
In the ARENA2 setting, two rays that allow interaction with both the world (i.e. 3D models) and the menu are attached to two motion-tracked gamepads, enabling two users to operate the system.
In the future, the HMD VR setting could also be adapted to support multiple players and a mode for remote HMD VR users to connect to a session in the ARENA2 is also possible.
One pair of tracked shutter glasses can be used to provide motion parallax for improved depth perception of the virtual world and if desired, active stereoscopy.
\Cref{fig:unreal} shows the default environment displayed to the user upon starting the application (\cref{fig:unrealstart}) as well as a screenshot from an HMD VR session with the menu widget opened (\cref{fig:unrealvr}).

\begin{figure*}[pos=htp]
    \centering
    \begin{minipage}[t]{.49\linewidth}
        \centering
        \adjustbox{valign=t}{\includegraphics[width=\linewidth]{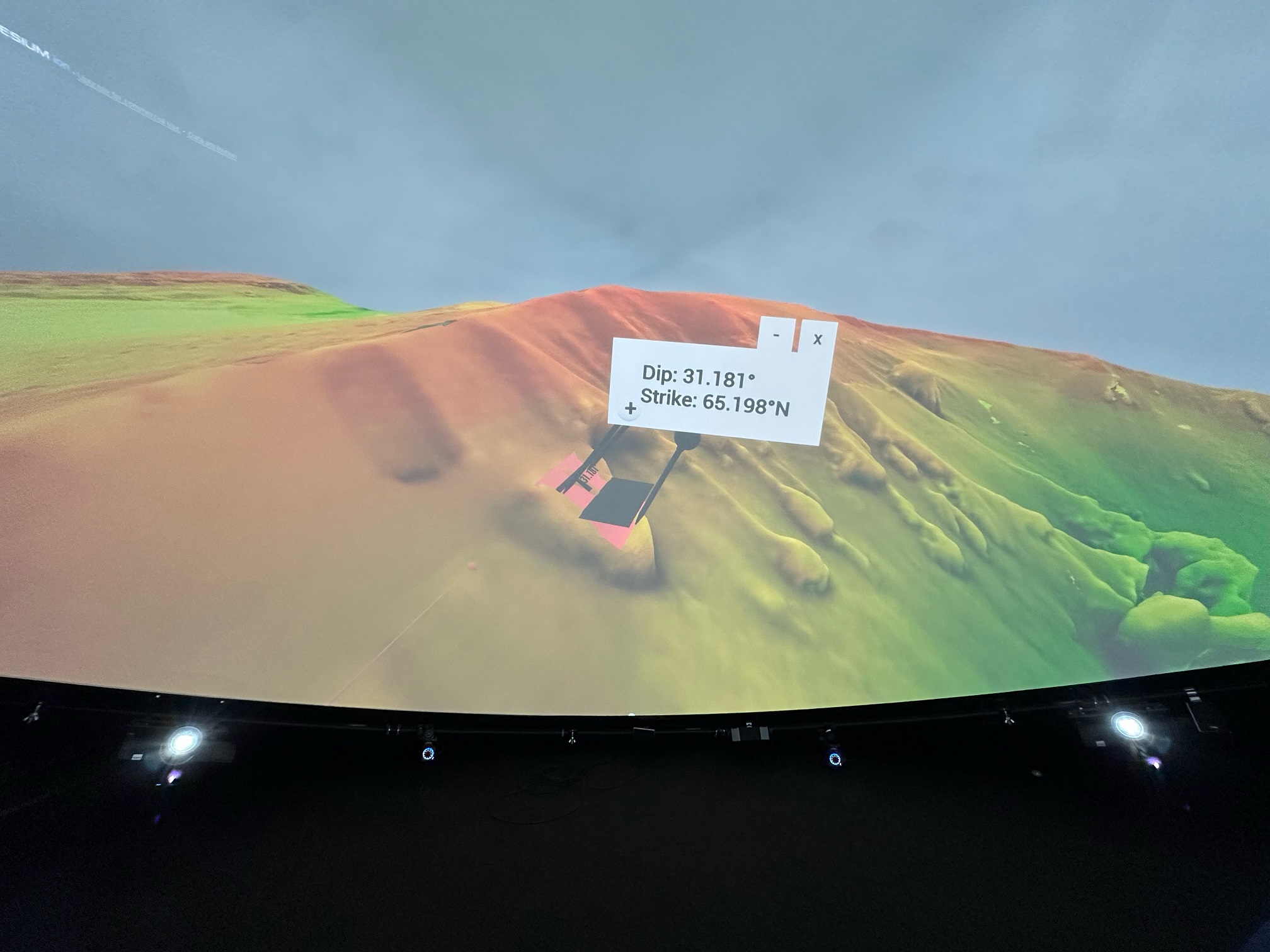}}
        \subcaption{}
        \label{fig:vf_arena1}
    \end{minipage}%
    \hfill
    \begin{minipage}[t]{.49\linewidth}
        \centering
        \adjustbox{valign=t}{\includegraphics[width=\linewidth]{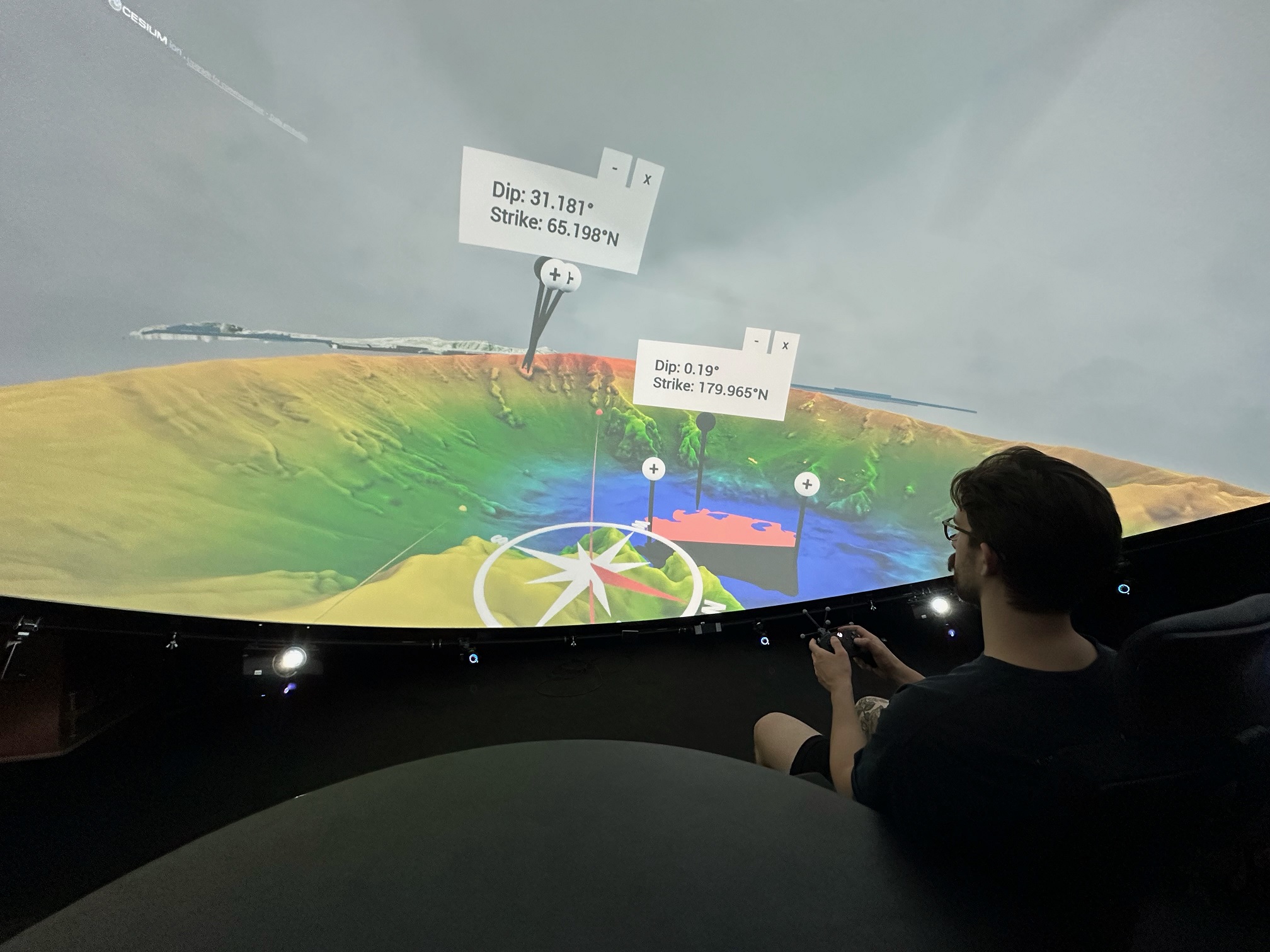}}
        \subcaption{}
        \label{fig:vf_arena2}
    \end{minipage}
    \vspace{1em} 

    \begin{minipage}[t]{\linewidth}
        \centering
        \adjustbox{valign=t}{\includegraphics[width=0.8\linewidth]{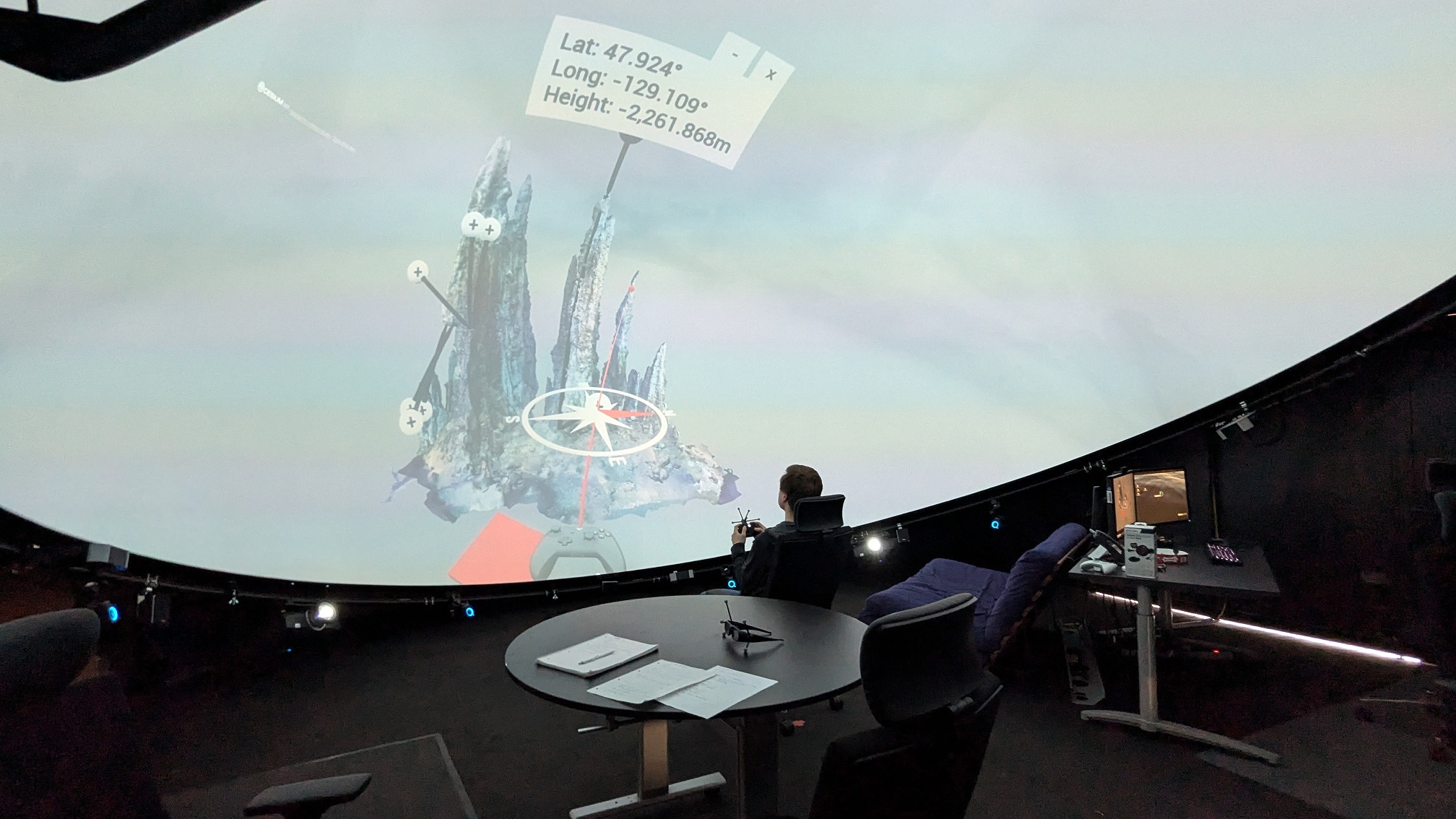}}
        \subcaption{}
        \label{fig:vf_arena3}
    \end{minipage}%

    \caption{Impressions of the VFT used inside the ARENA2. \Cref{fig:vf_arena1,fig:vf_arena2} show the Kolumbo dataset (see \Cref{subsec:usecase_kolumbo}) and \cref{fig:vf_arena3} shows the Mothra dataset (see \Cref{subsec:usecase_mothra}).}
    \label{fig:vf_arena}
\end{figure*}

\subsection{Interpretation Toolbox}
\label{subsec:vf-toolbox}

\begin{figure}[pos=h]
    \centering
    \begin{minipage}[t]{\linewidth}
        \centering
        \adjustbox{valign=t}{\includegraphics[width=\ifdim\linewidth<15cm\linewidth\else0.6\linewidth\fi]{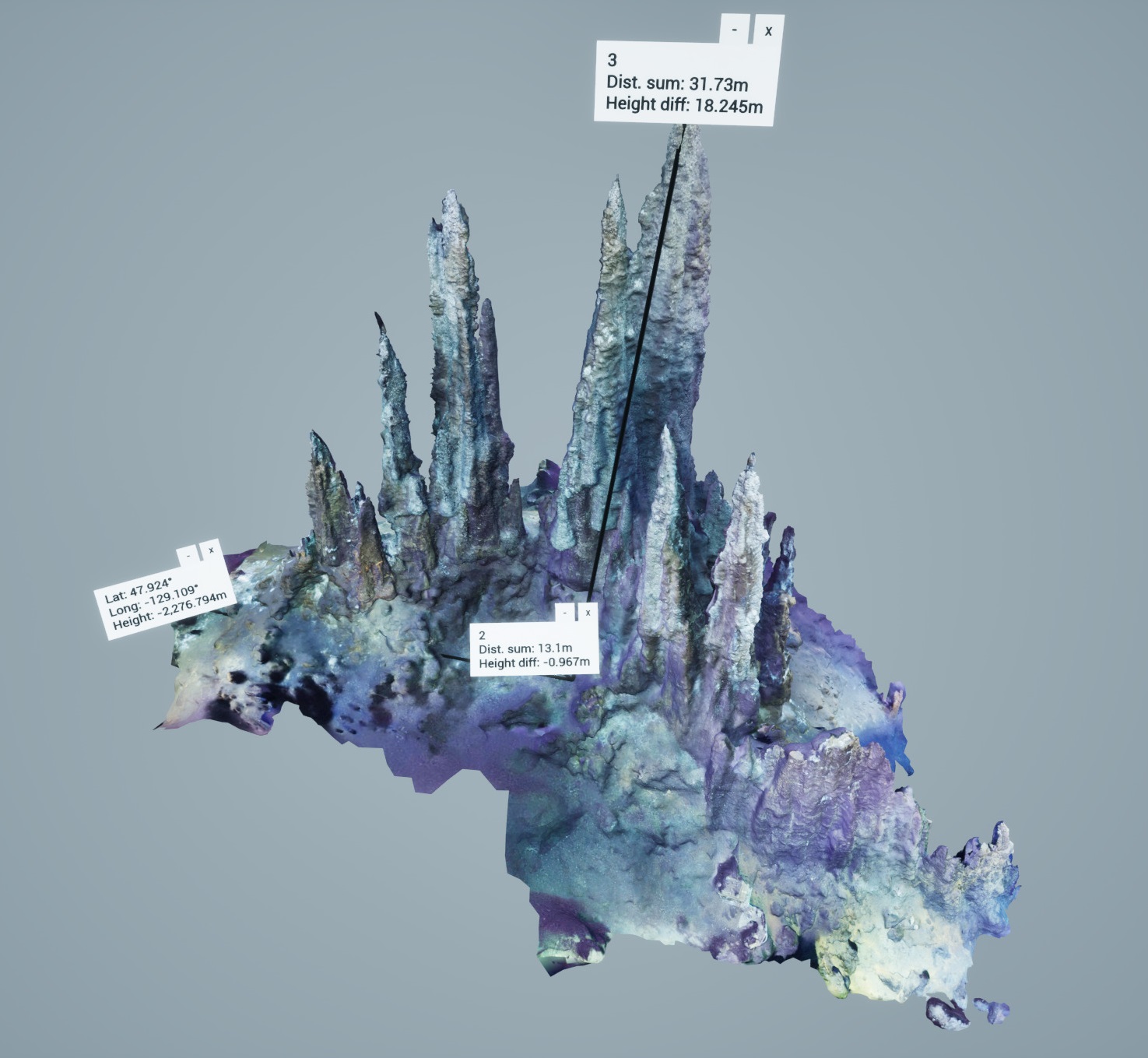}}
        \subcaption{Distance measurement along three markers shown on the Mothra model.}
        \label{fig:distance}
    \end{minipage}%

    \vspace{1em} 

    \begin{minipage}[t]{\linewidth}
        \centering
        \adjustbox{valign=t}{\includegraphics[width=\ifdim\linewidth<15cm\linewidth\else0.7\linewidth\fi]{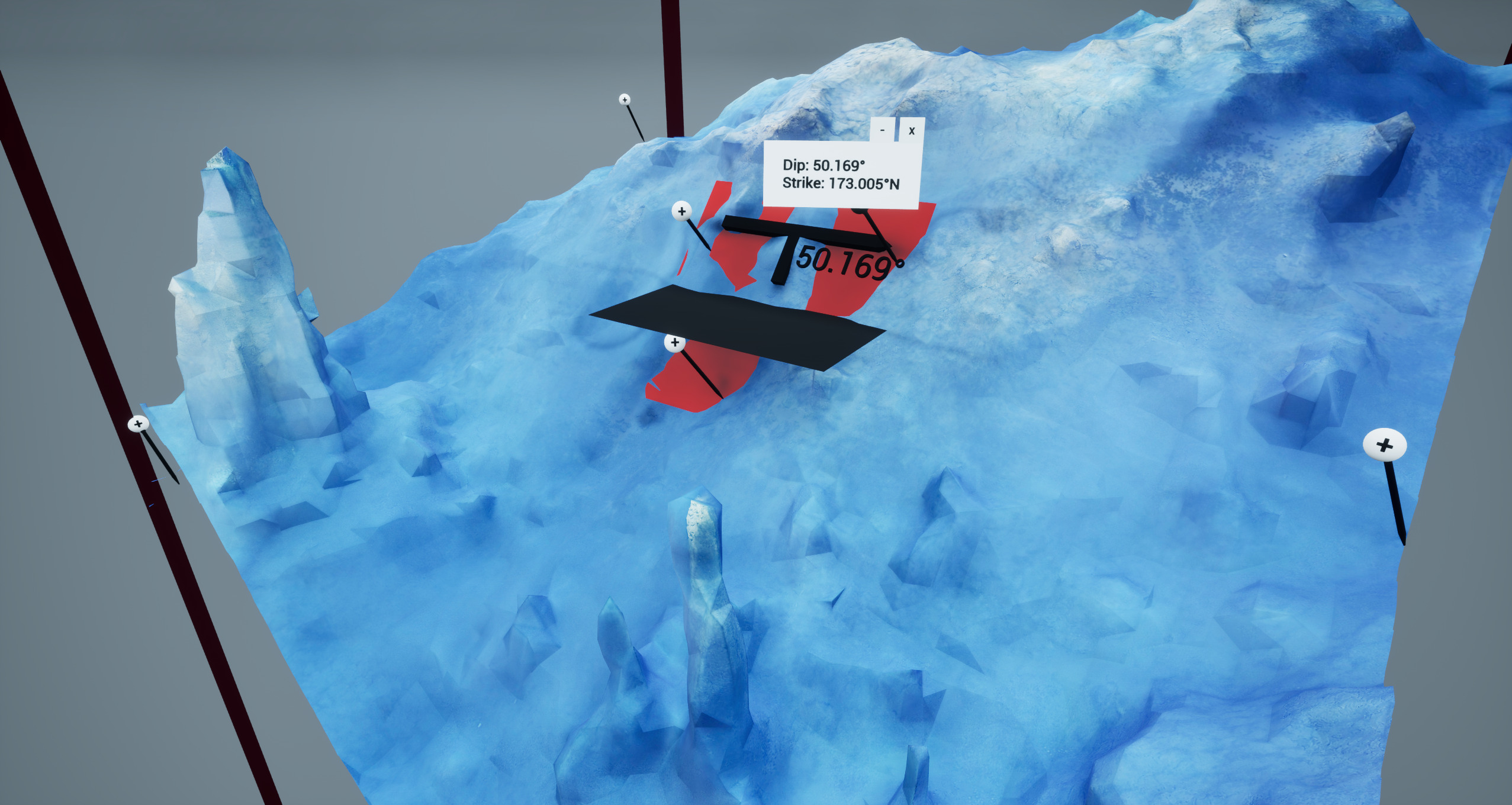}}
        \subcaption{Strike \& Dip measurement shown on a portion of the Niua South model, together with the vertical edges of a clipping box in the background. Two of the three markers that make up the strike plane have their labels hidden for visual clarity. This happens automatically when the measurement is generated but they can be set visible again by the user if desired.}
        \label{fig:strikedipclip}
    \end{minipage}%
    \caption{Screenshots from the Unreal Engine showing the results of measuring actions.}
    \label{fig:measurements}
\end{figure}

We devised an extensible collection of measurement options with the goal of providing both qualitative spatial insight and quantitative metadata as the outcome of geoscientific interpretation workflows.
The fundamental action upon which every measurement builds is the placement of a location marker to query the latitude, longitude, and elevation of that specific point.
These markers are visualized as pins with a label hovering above them. When a marker has been placed, the label initially shows the coordinates which are replaced by other data when the marker becomes part of a more complex measurement (see \cref{fig:unrealvr,fig:vf_arena}).

Any number from two to $N$ markers can be used to calculate a linear multi-segment distance measurement which reports the total distance along the sequence of markers, as well as the intermediate distances and elevation differences between two consecutive markers.
This feature is employed to measure arbitrary dimensions but also to characterize linear features such as faults, joints or artifacts of sediment transport.
\Cref{fig:distance,fig:kolumbounreal2,fig:niuaunreal1,fig:niuaunreal2} each show examples of the distance measurement.

\enquote{Strike and dip} are the format of geological notation for the azimuth and deviation from a leveled position in the orientation measurement of planar surfaces such as sedimentary bedding, faults, or rock faces.
This can be realized in our VFT by placing three markers that describe a plane in space.
Our VFT then derives and visualizes the angle between the horizontal plane and the strike plane i.e. the inclined surface. It also reports the maximum straight-line extent of the three-point measurement to give a measure of the sample scale.

A clipping box that renders the portion of a model outside its volume invisible can similarly be placed using at first two markers which define the box width and another defining the length. The height of the box is automatically set to encompass the entire elevation range of the cut-out part of the model.
\Cref{fig:strikedipclip} shows both a strike and dip measurement as well as a clipping box and \cref{fig:kolumbounreal2} also shows a clipping box.


To support quantitative studies beyond the visualization session, all measurement results can be exported to a JSON file. These files can also be loaded again from inside the VFT during runtime.

\section{Geoscientific Use Cases}
\label{sec:usecase}

In this section we introduce three use cases derived from recent studies of seafloor volcanology and hydrothermalism. Data are of varying scale (i.e. level of detail in overlapping sections) and heterogeneous origin, illustrating the range of likely situations in the real-world application of nested-scale seafloor surveying. Along these, we explore advantages and limitations of our VFT when applied to productive geoscientific work.
We aim to investigate whether differences in a) complexity or b) spatial context compared to real terrestrial fieldwork impact the acceptance and usage of the visualization by users.

\subsection{Use Case A: AUV Bathymetry of the Kolumbo Volcano, Santorini Greece}
\label{subsec:usecase_kolumbo}

\begin{figure}[pos=htp]
    \centering
    \begin{minipage}[t]{\linewidth}
        \centering
        \adjustbox{valign=b}{\includegraphics[width=\ifdim\linewidth<15cm\linewidth\else0.5\linewidth\fi]{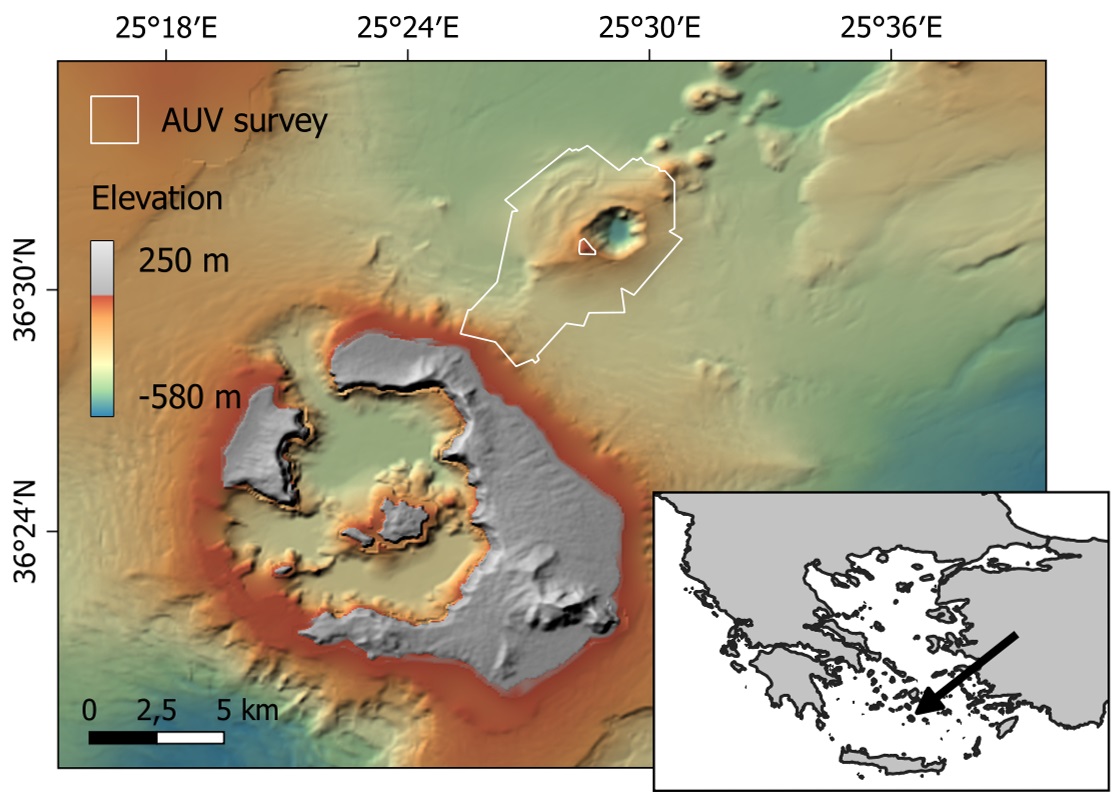}}
        \subcaption{}
        \label{fig:kolumbomap}
    \end{minipage}%
    \vspace{1em}
    \begin{minipage}[t]{\linewidth}
        \centering
        \adjustbox{valign=b}{\includegraphics[width=\ifdim\linewidth<15cm\linewidth\else0.5\linewidth\fi]{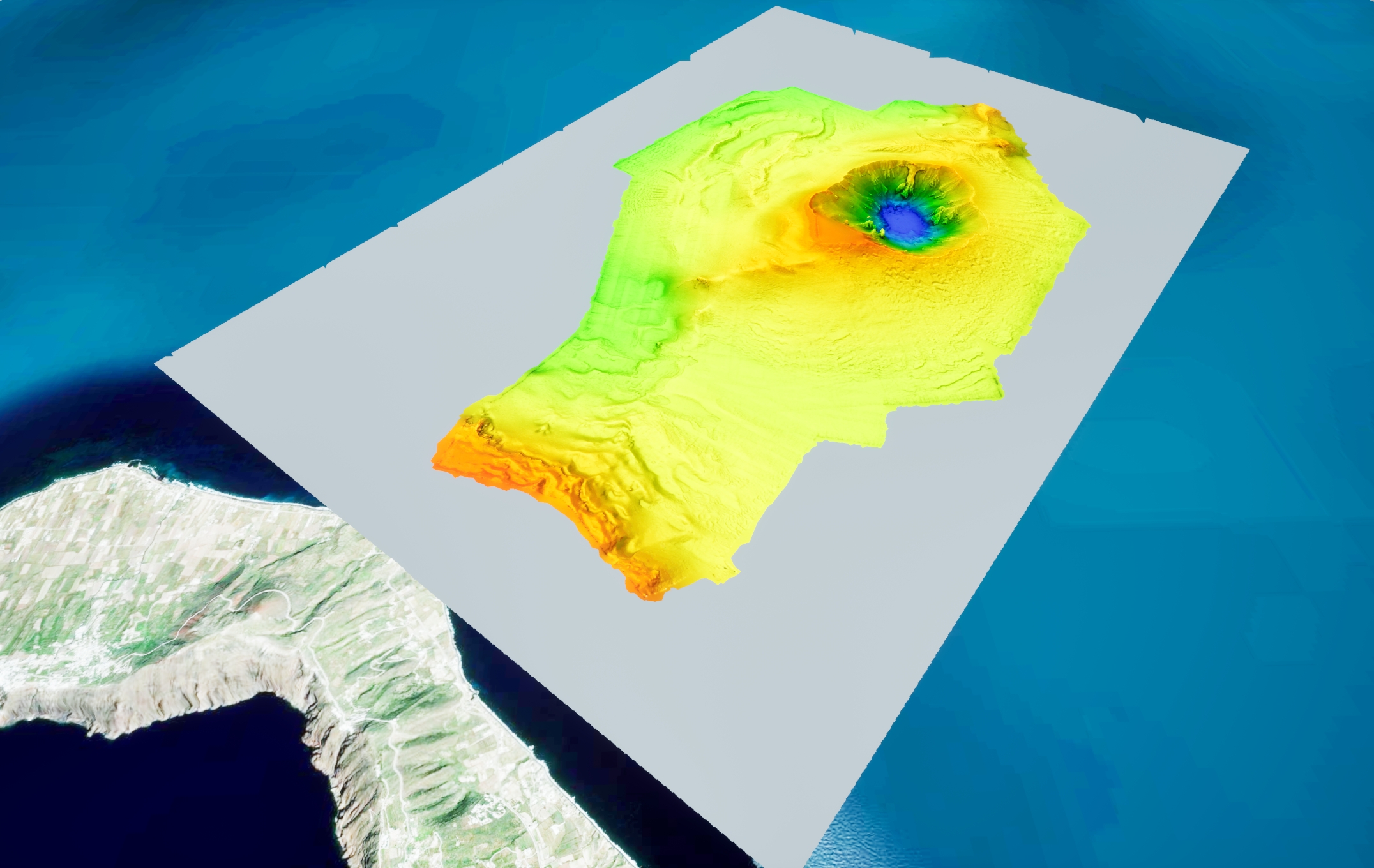}}
        \subcaption{}
        \label{fig:kolumbobird}
    \end{minipage}

    \vspace{1em} 

    \begin{minipage}[t]{\linewidth}
        \centering
        \adjustbox{valign=t}{\includegraphics[width=\ifdim\linewidth<15cm\linewidth\else0.5\linewidth\fi]{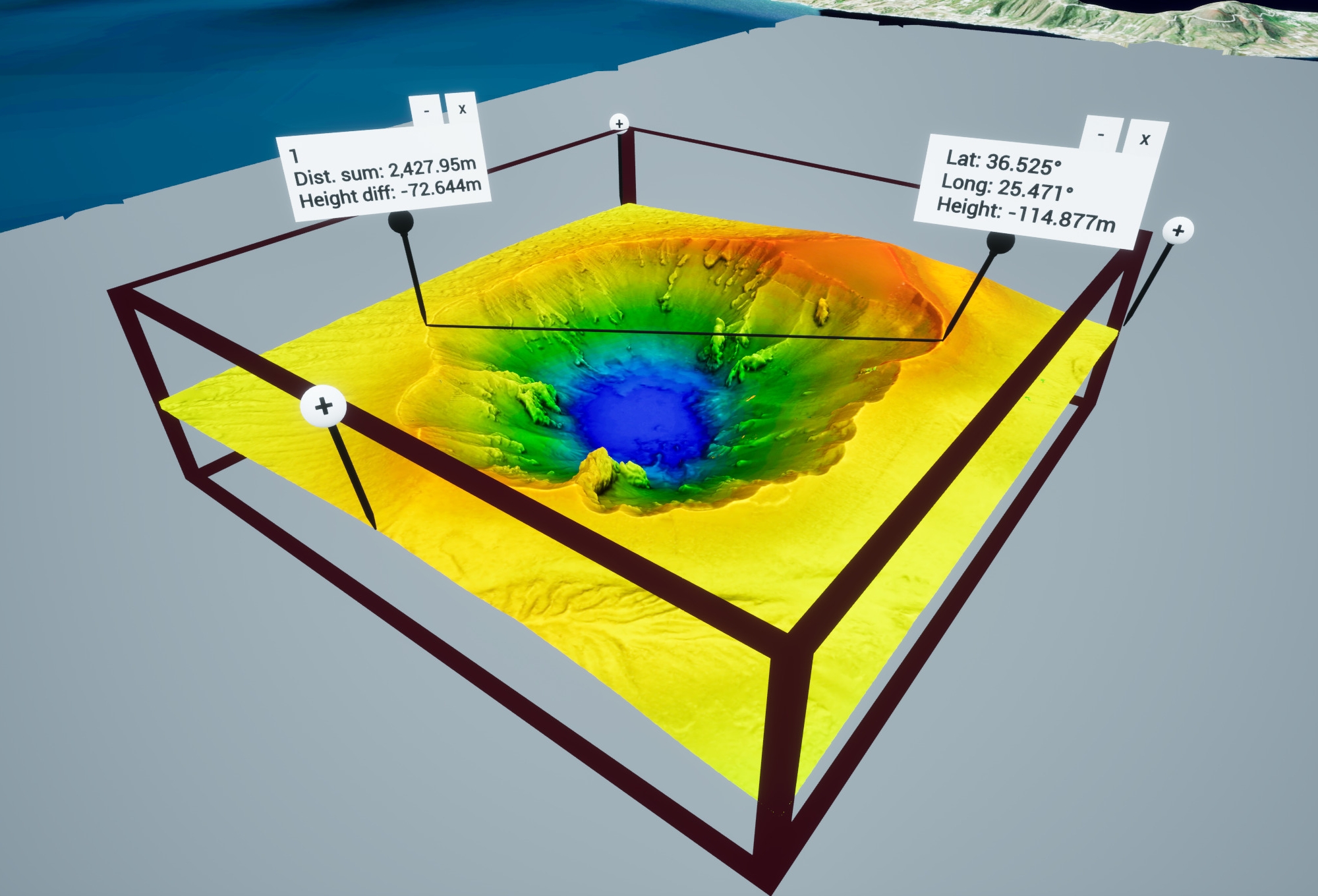}}
        \subcaption{}
        \label{fig:kolumbounreal2}
    \end{minipage}
    \caption[Caption for Kolumbo Figures]{Kolumbo is located near the island of Santorini in the greek Aegean Sea (\cref{fig:kolumbomap}).
        \Cref{fig:kolumbobird} shows a view of the Cesium 3D Tileset model in Unreal Engine (in desktop mode). In the lower left, Santorini is visible, visualized through the Cesium Bing Maps Aerial Imagery dataset (\url{https://cesium.com/platform/cesium-ion/content/bing-maps-imagery}).
        \Cref{fig:kolumbounreal2} shows a closer view of the crater in Unreal Engine. A distance measurement can be seen that measures the diameter of the crater, and a clipping box has been applied.}
    \label{fig:kolumbo}
\end{figure}

Kolumbo (\cref{fig:kolumbo}) is a submarine volcano located 7 km northeast of Santorini (Greece) in the Aegean Sea \citep{nomikou2012submarine}.
Kolumbo’s last eruption occurred in 1650 CE, when a slope instability triggered an explosion that formed a 500m deep and 2500m wide crater \citep{karstens2023cascading}.
Ongoing hydrothermal venting and seismicity confirm that Kolumbo is still active \citep{carey2013degassing,schmid2022heralds}. Moreover, a seismic full-waveform inversion indicates the presence of a shallow magma reservoir about 2km beneath the seafloor \citep{chrapkiewicz2022magma}.
Our bathymetric dataset has a horizontal resolution of 2m, an extent of 10 km by 6 km and was acquired in 2017 during research cruise POS510 onboard RV Poseidon \citep{2018poseidon} by the GEOMAR AUV ABYSS equipped with a RESON Seabat 7125 multibeam echo sounder. The dataset is available at the PANGAEA data repository \citep{petersen2023mbpd}.

The prominent morphological topics to be explored during virtual fieldwork are the structural context and immediate contacts of several exposed dykes within the inner crater wall as well as the extent of ubiquitous late-stage lava flows down the crater slopes. The strike and dip as well as the distance measurements further allow the quantitative assessment of slope segments at the over-steepened and internally deformed northwestern flanks in comparison to the intact southeastern sector.

\subsection{Use Case B: ROV based Photogrammetric Model at Mothra Hydrothermal Field, Endeavour Ridge}
\label{subsec:usecase_niuasouth}

\begin{figure}[pos=ht]
    \centering
    \begin{minipage}[b]{\linewidth}
        \centering
        \adjustbox{valign=t}{\includegraphics[width=\ifdim\linewidth<15cm\linewidth\else0.5\linewidth\fi]{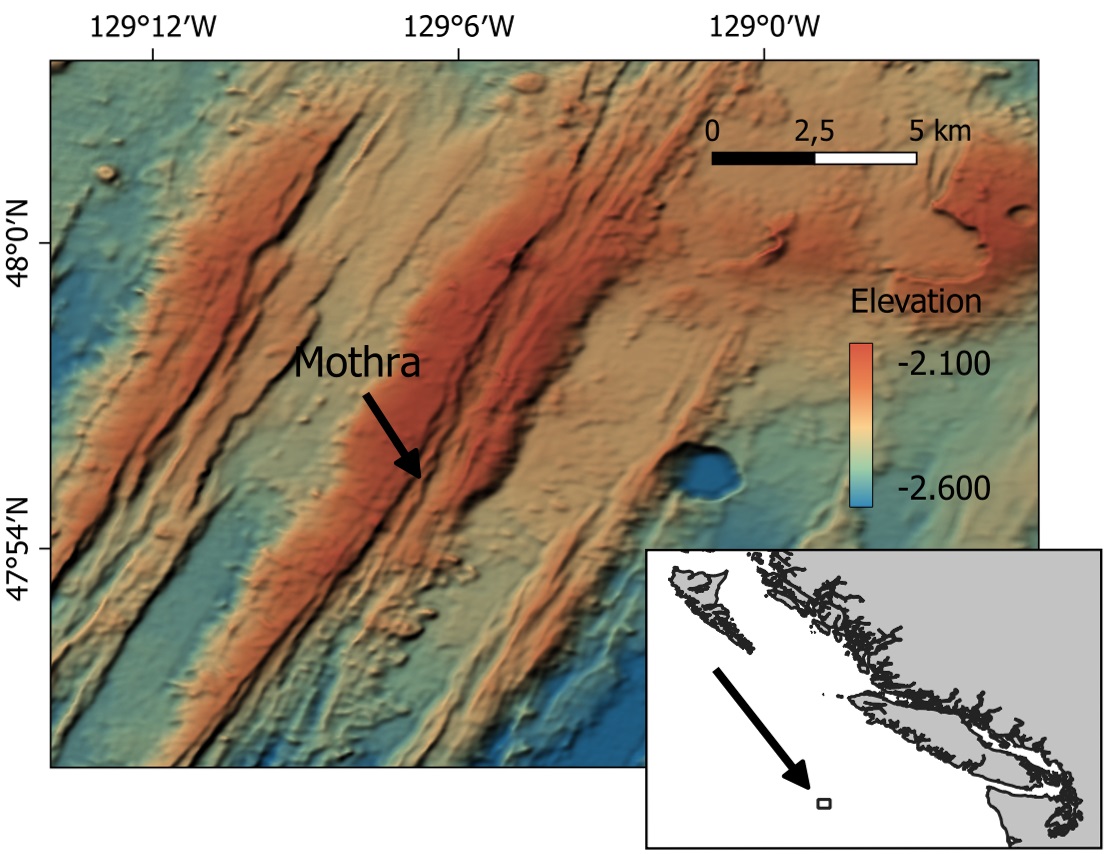}}
        \subcaption{}
        \label{fig:mothramap}
    \end{minipage}%
    \hfill
    \begin{minipage}[b]{\linewidth}
        \centering
        \adjustbox{valign=t}{\includegraphics[width=\ifdim\linewidth<15cm\linewidth\else0.5\linewidth\fi]{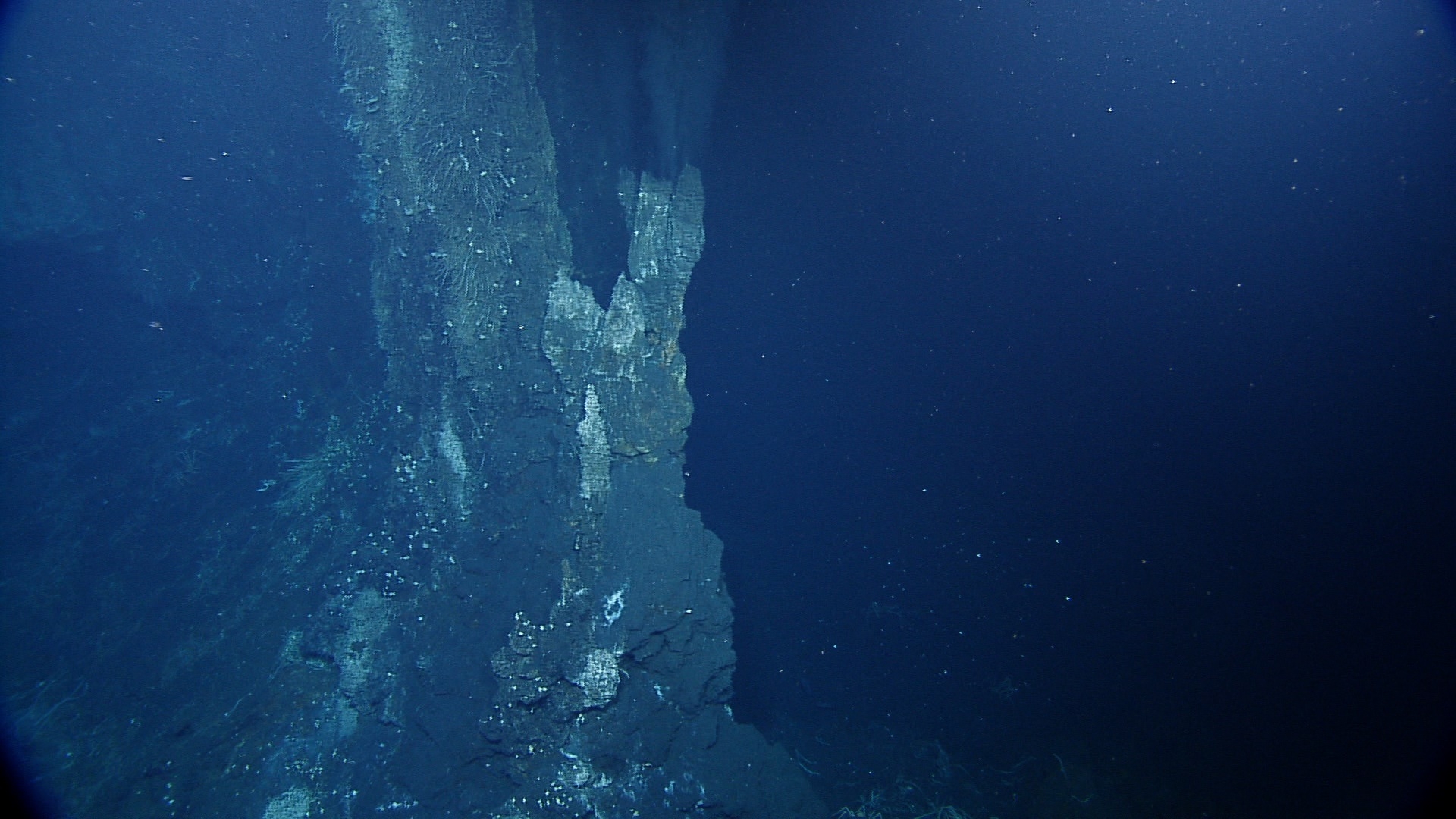}}
        \subcaption{}
        \label{fig:mothrafoto}
    \end{minipage}

    \caption{The Mothra hydrothermal field is located at the Endeavour Segment of the Juan de Fuca Ridge, NE Pacific (\cref{fig:mothramap}). \Cref{fig:mothrafoto} shows one shot from the footage taken for photogrammetric reconstruction. Compare \cref{fig:distance} for a view of the full model.}
    \label{fig:mothra}
\end{figure}

Mothra (\cref{fig:mothra}) is the southernmost of five major hydrothermal fields at the Endeavour Segment of the Juan de Fuca Ridge, NE Pacific, which have been regularly visited for the last 30 years \citep{delaney1992geology,clague2020hydrothermal}. This means these are not only one of the earliest discoveries of seafloor hydrothermalism but also that they count among the best studied seafloor outcrops in general. Faulty towers is one of the largest of six hydrothermal complexes at Mothra Field, measuring roughly $30\times 13 \times 20$ meters hosting  a dozen slender, diffusely venting sulfide spires with minor occurrences of black smoker venting towards the perimeter of the group. The chimneys are roughly aligned along a N-S striking fault related to the local tectonics along the axial valley \citep{robigou1993large,kelley2001geology,glickson2007geology}. During the 2015 Ocean Networks Canada maintenance cruise NA069, we conducted a photogrammetric survey of the entire complex using the ROV HERCULES (on dive H1960) and its ZeusPlus HD camera. The resulting model has a spatial resolution ranging between 5 and 15cm with a textural resolution locally down to $<1$ cm.

\Cref{fig:mothrafoto} shows a shot from the footage recorded for photogrammetric reconstruction. Noticeable here is that without further processing (e.g. stitching photos together in a photomosaic, or photogrammetric reconstruction) it is difficult to grasp the full extent of the surroundings and discern relevant details from a single photo.

Due to its resolution, color texture and steep relief, this dataset is well suited for a number of digital outcrop studies focusing on habitat mapping of the vent fauna, structural measurements on spacing, vertical extent, and alignment of chimney edifices to the underlying fluid-feeding fault system, as well as the quantification of the talus on which part of the constructive edifice resides.

\subsection{Use Case C: Nested Photogrammetric and Hydroacoustic Bathymetry at Niua South Vent Field, Tonga, SW Pacific}
\label{subsec:usecase_mothra}

\begin{figure*}
    \centering
    \begin{minipage}[t]{.49\linewidth}
        \centering
        \adjustbox{valign=b}{\includegraphics[width=\linewidth]{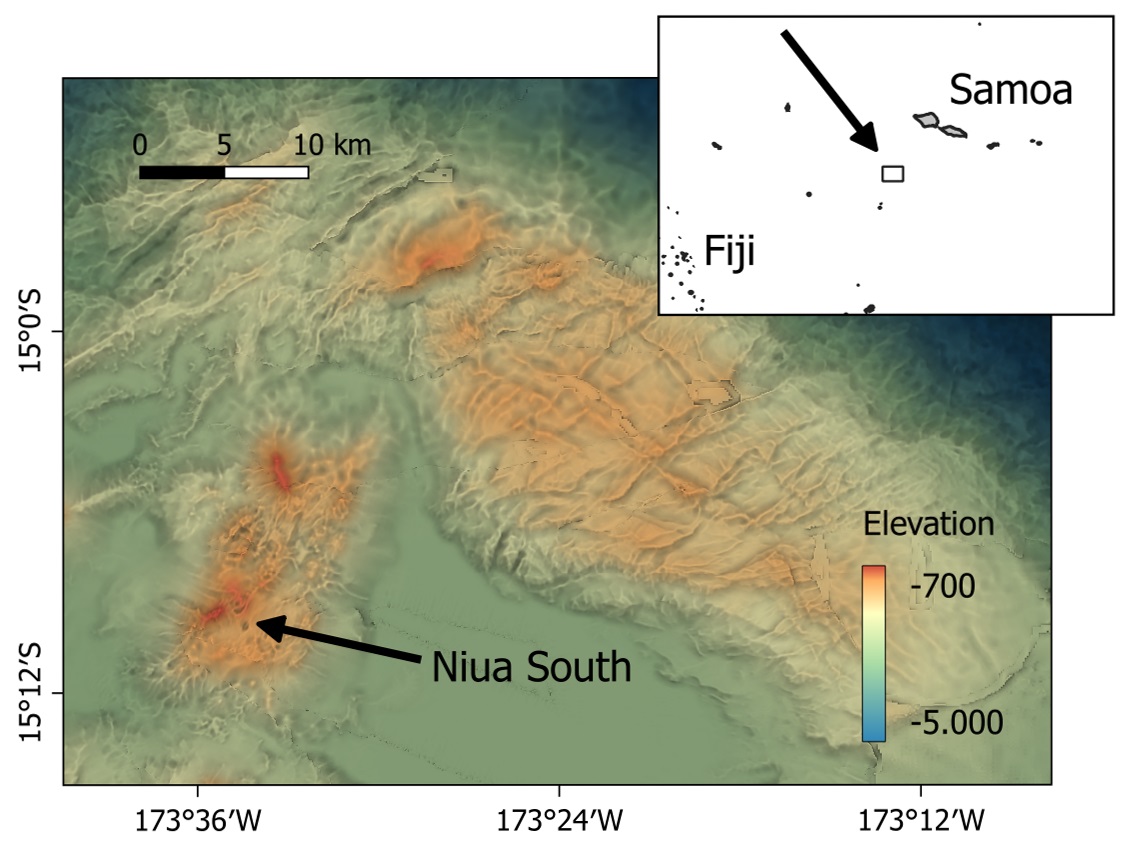}}
        \subcaption{}
        \label{fig:niua_map}
    \end{minipage}%
    \hfill
    \begin{minipage}[t]{.49\linewidth}
        \centering
        \adjustbox{valign=b}{\includegraphics[width=\linewidth]{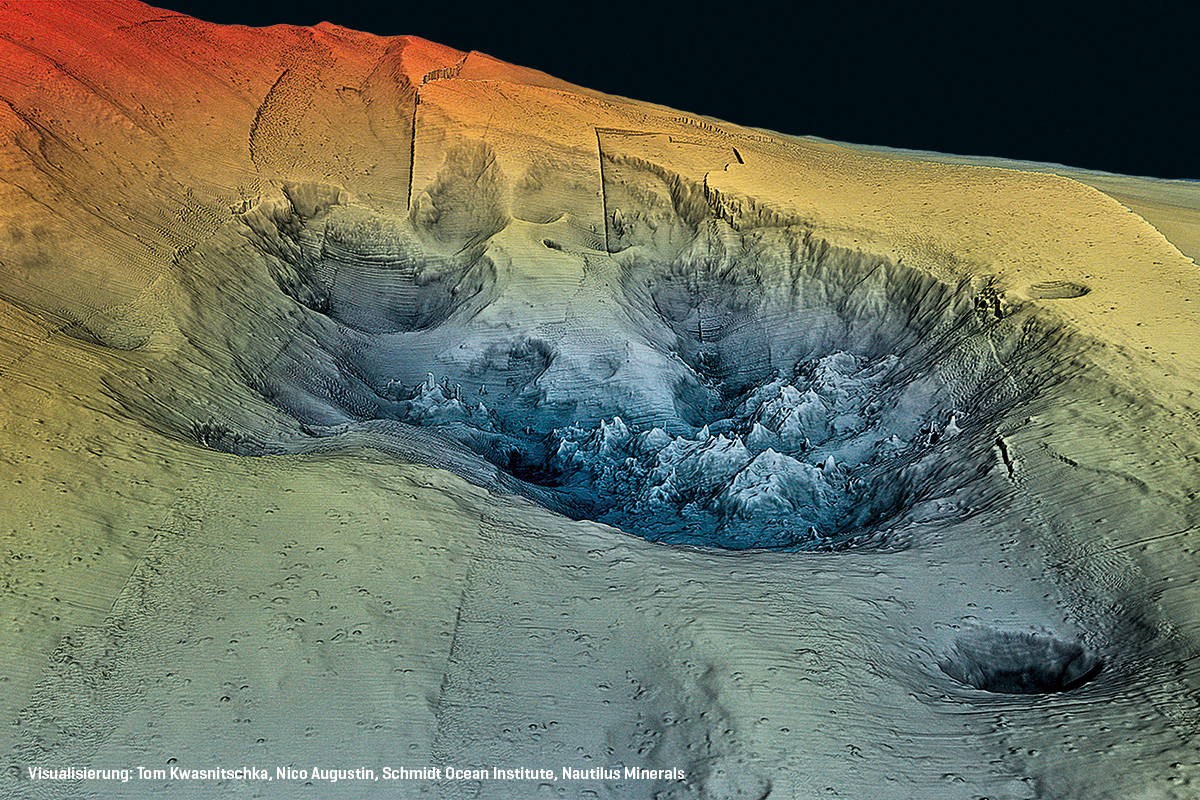}}
        \subcaption{}
        \label{fig:niuasouth2}
    \end{minipage}

    \vspace{1em} 

    \begin{minipage}[t]{.49\linewidth}
        \centering
        \adjustbox{valign=t}{\includegraphics[width=\linewidth]{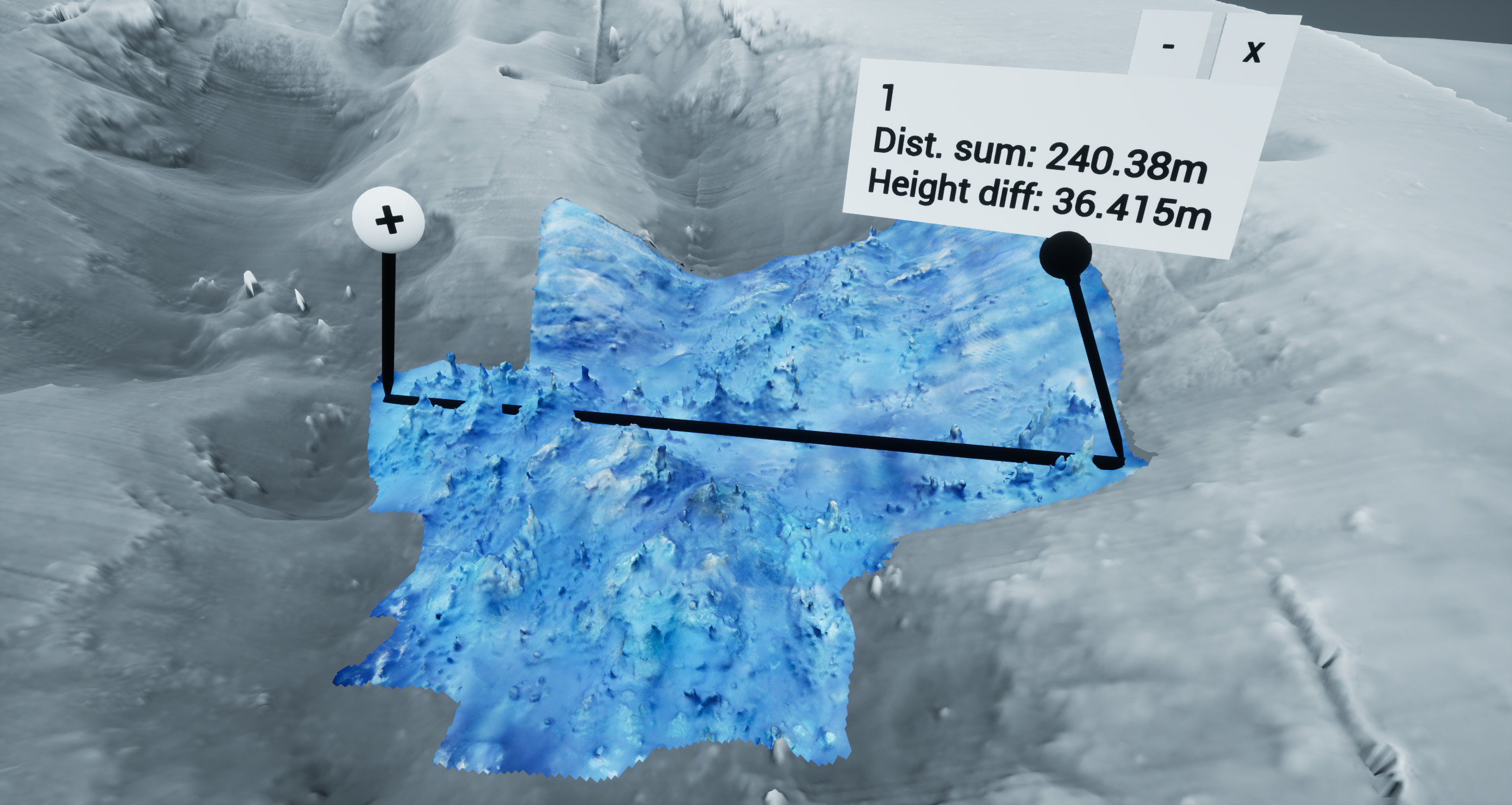}}
        \subcaption{}
        \label{fig:niuaunreal1}
    \end{minipage}%
    \hfill
    \begin{minipage}[t]{.49\linewidth}
        \centering
        \adjustbox{valign=t}{\includegraphics[width=\linewidth]{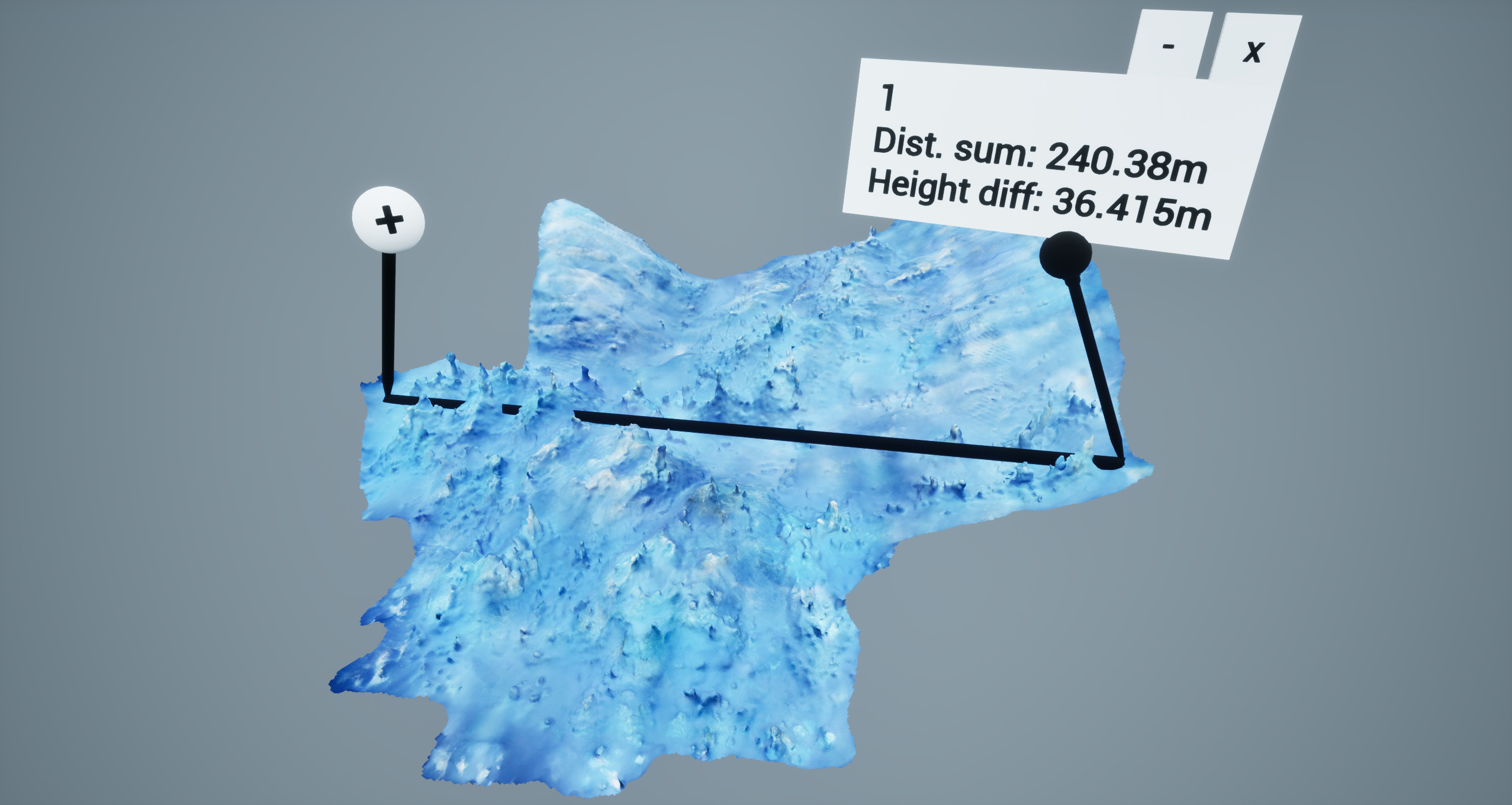}}
        \subcaption{}
        \label{fig:niuaunreal2}
    \end{minipage}
    \caption{The Niua South volcano is located in the Lau basin between
        Fiji and Samoa (\cref{fig:niua_map}).
        \Cref{fig:niuasouth2} shows a vertically exaggerated digital elevation model of the Niua South volcano.
        \Cref{fig:niuaunreal1,fig:niuaunreal2} show the central part of the Niua South hydrothermal field model in Unreal Engine with and without its surrounding area which is lower in resolution. A distance measurement shows the distance across the central part of the model in meters.}
    \label{fig:niuasouth}
\end{figure*}

Niua South (\cref{fig:niuasouth}) is one of two hydrothermal fields near the summit of Niua Volcano at the northern terminus of the Tonga volcanic arc. It is situated within a 500m wide crater at a depth of 1180m. Within a central area of 150 by 200 m, several dozen active hydrothermal chimneys of up to a height of 8m rise from sulphide talus mounds up to 10m high. Spread across these mounds are over a hundred of >1m tall, inactive and partly collapsed chimneys \citep{peters2021disproportionation,gartman2017boilinginduced}. The site was commercially surveyed in 2011 using the GEOMAR AUV ABYSS and bespoke RESON Seabat 7125 multibeam echo sounder yielding a 2m resolution hydroacoustic map of the entire crater and its surroundings measuring 2 km across. In 2016, we surveyed it again using the ROV ROPOS aboard R/V Falkor (cruise FK160320) \citep{kwasnitschka2016virtual}, yielding a dataset of 229,000 images. For the purpose of this study, we employed a subset of this data covering the central 200m by 200m which yielded a preliminary photogrammetric model at an average geometrical resolution of 15cm and a textural resolution of locally up to 5 cm. The two datasets were superimposed on each other within an estimated error of < 2.5m laterally based on fiducial landmarks, which is sufficient within the scope of our study.

The resulting synthesized terrain model is perfectly suited for habitat assessments as well as the study of abundance, location and physical properties (height, diameter, orientation) of sulfide chimneys, and to map the extent of talus mounds.
It uniquely allows the synoptic observation of morphological phenomena across an entire vent field, including the far field structural context of its surroundings.
The fact that the vent field lies surrounded by crater walls implies that the effect of immersion into the scene may not be corrupted by the visibility of the boundary of the model, when viewed in a spatially immersive visualization environment.


\section{Discussion}
\label{sec:discussion}
The availability of high-resolution bathymetry data is continuously increasing in light of endeavors such as Seabed 2030 \citep{mayer2018nippon}.
While this increasing amount of data is also being analyzed using machine learning approaches, it relies on human classification efforts and thus the personal observation and analysis by researchers is not becoming obsolete. Moreover, the incorporation of strategic ground truthing of regional observations against local human-scale outcrops at the seafloor becomes ever more important to validate inferences on larger scales.
Therefore, offering researchers an alternative tool to analyze seafloor models in a more immersive way than on PC monitors may yield advantages both in productivity and in the quality of insight.

In order to gain a first-order, qualitative round of feedback on our development, we invited seven domain scientists, who are not involved in the project, in varying group sizes (from 1-3) to explore our tool both in the ARENA2 and with HMD VR. None had prior experience with using our tool, however, several had previously been in the ARENA2 or had worn VR headsets.
This informal gathering of feedback was done as a preliminary step towards a semi-structured qualitative HCI evaluation study planned for future publication.

All users were first shown the ARENA2 setting, familiarized with the controls, menu navigation and measurement tools, and encouraged to explore all three datasets and apply all available measurement options.
Afterwards they were also shown the HMD VR version.
They provided anecdotal feedback and suggestions for improvements which we discuss below together with our own observations. We omit concrete suggestions regarding the implementation (e.g. movement, controls, visual design decisions) from this discussion.

\subsection{General Results}

An important aspect we are facing occasionally when working with our ARENA2 laboratory is the additional effort that has to be made for researchers to visit the laboratory which impacts the overall value of working with the VFT.
This value of a visualization is defined by \cite{vanwijk2005value} in the following way:

\begin{quote}
    \enquote{[...] a great visualization method is used by \textbf{many people}, who use it \textbf{routinely} to obtain \textbf{highly valuable knowledge}, without having to \textbf{spend time} and \textbf{money} on hardware, software, and effort.}
\end{quote}

The obtaining of knowledge is specified by \cite{vanwijk2005value} as $\Delta K$, the difference between prior knowledge and the increased knowledge after interacting with a visualization.

In the following, we make callbacks to this definition by highlighting whether or not these points hold true for the VFT. The aforementioned effort of visiting the laboratory for example increases the \textbf{time spent on effort}.

Our vision for the VFT prototype was to provide a tool for scientific sensemaking and analysis for researchers (thus providing \textbf{highly valuable knowledge}).
However, feedback by our testing users often mentioned the potential for the tool to rather be used in a more educational context and for outreach, additional to scientific work, an approach \cite{ynnerman2018exploranation} call exploranation (a portmanteau of exploration and explanation).
This could be implemented by taking an approach more in the line of \cite{zhao2019harnessing}, \cite{klippel2019transforming,klippel2020value} and also \cite{tibaldi2020worldbased}.
Their approaches for example show text-based information, photographs and figures about the visualized 3D model, and offer hints for their users which are mostly meant to be geoscience students using the application for training.
Enhancing the tool with external information and additional data might provide increased value by increasing the number of people that are using it ($\Rightarrow$ \textbf{many people}).

Generally, users mentioned that the application has too little data variability
with there being only 3D models available.
One user remarked that it is \enquote{important for [them] to have the big picture}, e.g. include high resolution photogrammetry embedded in a surrounding context of lower resolution bathymetry (see \cref{fig:niuaunreal1,fig:niuaunreal2}).
There is much data available, often from the same location so it is important to combine those to have a more complete context. This confirms our assumption that lack of either complexity or spatial context compared to real terrestrial fieldwork may lead to a rejection of the presentation by the users.

Our assumption in accord with related literature (see \cref{subsec:bg-virtualfieldwork}) is that 3D models help with understanding structures from a different perspective compared to 2D on a desktop. Working with highly trained professionals, however, the scientific objective of our testing users was found to be on quantitative measurements and less on construction or refinement of a mental model through visual data exploration.
Our users deemed the former to still be more productive on a flat PC monitor with the latter possibly providing some inspiration beforehand. Therefore in this scenario $\Delta K$ would be less in the ARENA2 for experts with a large amount of prior knowledge about the data.
A quote regarding this was \enquote{Why would it be necessary to be in here and not work on this on a PC?} which also touches the aspect of users being reluctant to \textbf{routinely} use the VFT if their $\Delta K$ is not perceived as sufficient and they have to invest \textbf{effort} into visiting the ARENA2.

For scientific analysis and sensemaking, the VFT needs to implement the standard tools, that are used on PC, in a way so that they give additional value in the immersive setting and are intuitive for geoscientists (thus less \textbf{time spent} on learning).
The Strike and Dip tool for example was criticized for not being able to measure on a small enough scale.

Still though, one user commented that they \enquote{got a different perspective on the model} and during the session remarked that they now suspected an opinion they had held about a dyke in the Kolumbo crater might have to be reevaluated ($\Rightarrow$ \textbf{highly valuable knowledge}/$\Delta K$).
While the approach of being a geoscientific interpretation tool is important, users remarked that they see our VFT at an earlier point in the workflow and as having a different strength. Namely the ability to visualize 3D models in a different way than a PC monitor which opens a new basis for discussion that leads to idea generation for more precise interpretation.

The topic of \textbf{money spent} could be touched here with regards to the general inaccessiblity of a facility such as the ARENA2. Our application is, however, usable both with VR head sets which - depending on model - are relatively affordable, as well as on a desktop PC although the latter does lack the immersive aspect.

One consensus between our test users was that a clear benefit of using the VFT in the ARENA2 is the possibility for collaborative group work and discussion instead of working on one's own ($\Rightarrow$ \textbf{many people}).
Testing users suggested the tool could be used to either discuss initial results after a research cruise or discuss an upcoming follow-up cruise and figure out the most important and promising locations to (re-) visit and investigate.
The HMD VR setting in its current state was generally not considered an alternative to productive work on a desktop PC, lacking both the extended toolset of desktop software as well as the option of collaborative work and easily being able to take notes in the ARENA2.

\subsection{Dataset Selection and Presentation}

Most comments on the visualized data were in regards to pre-processing topics such as the chosen color map and shading of the model or more generally the resolution.

The Mothra photogrammetry model was liked both in the ARENA2 and VR due to its high resolution which according to user comments \enquote{almost [felt] like you can touch it} which according to them improved their feeling of immersion.
Nevertheless, the surrounding area in which the model could be embedded is missing and therefore not much of the larger context was able to be used for sensemaking.
The missing surrounding spatial context also caused reports of disorientation and lack of a motion reference.

Large datasets, such as the Kolumbo model where the bathymetry is being investigated, were found to be more suited to the ARENA2 despite also being criticized as feeling like \enquote{it's just a bigger screen than PC}.
In contrast, small scale, high-resolution photogrammetry datasets (Mothra), where features as small as crabs are visible and measurable, were deemed more fitting for HMD VR.

~\\
Overall, our prototype was commended for its novel (from the stand point of our test cohorts) approach and immersive, interactive visualization which none of our domain expert testing users had worked with previously.
Of course, an approach like this will not be able to compete with desktop-based GIS software in terms of the sheer number and variety of tools available for productive scientific work but it can help form opinions and get a different perspective on the visualized datasets.
By broadening the niche of our prototype to think of it as more than a tool for scientific sensemaking and include aspects of education and outreach, it might prove to be a valuable addition to scientific work at our ocean research institute and potentially elsewhere.
Therefore, referring back to \citeauthor{vanwijk2005value}, our tool has the potential to be used by \textbf{many people} to obtain \textbf{highly valuable knowledge} thus increasing their $\Delta K$ although they might have to \textbf{spend time} especially if it were to be used \textbf{routinely}.


\section{Conclusion}
\label{sec:conclusion}
Many classically trained geologists and other geoscientists make use of their spatial visualization skills honed in fieldwork to make sense of 3D geospatial data.
Doing this in immersive environments has shown to provide new perspectives on the data especially if it's data from locations that cannot easily be visited in person or put in realistic perspective on a simple computer screen.
Implementing an immersive application for virtual fieldwork will not make obsolete the extensive toolboxes that established GIS softwares offer.
It can, however, give scientists a new way to interact with data.
Implementing such an application in the framework of a freely available and popular game engine is an approach to making the development sustainable and future-proof for extension.

The application introduced here is built for a unique spatially immersive projection dome to specifically enhance the scientific work of researchers at our institute, though it also supports HMD VR to offer users outside our very specialized environment to explore the tool.
Our initial datasets included in our prototype are real-world use cases that are being investigated by multiple working groups at GEOMAR.
We visualize these data in our prototype to be explored and examined using quantitative measurement tools in VR and collaboratively in the ARENA2.

The development of such software is difficult to call \enquote{finished} at any point. Therefore the future work on our tool is manifold: (a) implement more measurement options such as rake measurement for fault kinematics, height profile, volume and area measurement, (b) include more datasets and develop a parameterized import functionality at runtime, (c) connect with a history management companion application \citep{bernstetter2023practical}, and of course (d) improve overall design and usability.

\printcredits

\section*{Declaration of competing interest}
The authors declare that they have no known competing financial interests or personal relationships that could have appeared to influence the work reported in this paper.

\section*{Acknowledgements}
\label{sec:ack}
Armin Bernstetter is a doctoral researcher funded through the Helmholtz School for Marine Data Science (MarDATA), Grant No. HIDSS-0005.


\section*{Code availability section}

\begin{itemize}
    \item Name of the code/library: Virtual Fieldwork Unreal Engine Application
    \item Contact: abernstetter@geomar.de
    \item Hardware requirements: Requirements for Unreal Engine 5.3.2
    \item Program language: C++, Unreal Engine Blueprint Visual Scripting
    \item Software required: Unreal Engine 5.3.2
    \item Program size: Repository size ca. 800mb, size on disk after installation of dependencies and compilation 5-10gb
    \item The source code is available for downloading at the link: \url{https://git.geomar.de/arena/unreal-development/virtualfieldwork}
    \item Executable: An executable file/packaged Unreal Engine application can be requested by contacting the author

\end{itemize}
\section*{Data availability section}
The data used in this paper is partially available.

The data from which the Kolumbo Cesium3DTileset was created is available via: \url{https://doi.pangaea.de/10.1594/PANGAEA.958275}

We have not yet made the Cesium3DTilesets of our use cases publically available.

\section*{Declaration of generative AI and AI-assisted technologies in the writing process}
Statement: During the preparation of this work the authors used ChatGPT in order to check grammar and punctuation, find synonyms, and rephrase sentences to avoid repetitive text or redundant formulations. After using this tool, the authors reviewed and edited the content as needed and take full responsibility for the content of the publication. All content, interpretations, conclusions, and any remaining errors are our own.

\bibliographystyle{cas-model2-names}
\bibliography{references}

\end{document}